\documentclass{aa}
\usepackage{txfonts}
\usepackage{float}
\usepackage{graphicx}
\usepackage{subfigure}
\usepackage{tabularx}
\usepackage{natbib}
\usepackage{comment}
\usepackage{color}
\def\cha {\textit{Chandra}}

\def\arcsec{\hbox{$^{\prime\prime}$}}

\def\xmm{\textit{XMM-Newton}\/}

\def\arcsec{\hbox{$^{\prime\prime}$}}
\def\z{{\it z\/}}
\def\Ms{M$_{\odot}$\/}
\def\ecmss{erg\,s$^{-1}$\,cm$^{-2}$}
\def\es{erg\,s$^{-1}$}

\usepackage{multirow}
\usepackage{makecell}

\begin{document}
\title{The active CGCG 077-102 NED02 galaxy within the Abell 2063 galaxy cluster
 ~\thanks{Mainly based on observations obtained with \textit{Chandra}, SDSS, and OHP observatories 
(see acknowledgements for more details).}}

\author{
C.~Adami\inst{1} \and
K.~Parra Ramos\inst{1,3} \and
J.T.~Harry  \inst{2}  \and
M.P.~Ulmer\inst{2} \and
G.B.~Lima Neto \inst{3} \and
P.~Amram  \inst{1} 
}

\offprints{C. Adami \email{christophe.adami@lam.fr}}

\institute{
Aix Marseille Univ, CNRS, CNES, LAM, Marseille, France
\and
Department of Physics and Astronomy and CIERA, Northwestern University, 2145 Sheridan Road, Evanston, IL 60208-3112, USA
\and
Instituto de Astronomia, Geof\'{\i}sica e Ci\^encias Atmosf\'ericas, Universidade de S\~ao Paulo, Rua do Mat\~ao 1226, 05508-090 S\~ao Paulo/SP, Brazil
}

\date{Accepted . Received ; Draft printed: \today}

\authorrunning{Adami et al.}

\titlerunning{The active CGCG 077-102 NED02 galaxy within the Abell 2063 galaxy cluster}

\abstract 
{Within the framework of investigating the link between central super massive black holes
in the core of galaxies and the galaxies themselves, we detected a variable X-ray source
in the center of CGCG 077-102 NED02, member of the CGCG 077-102 galaxy pair within the Abell 2063 cluster of galaxies.}
{Our goal was then to combine X-ray and optical data to demonstrate that this object harbors an active super massive black hole in its core, and to relate this to the dynamical status of the galaxy pair within the Abell 2063 cluster.}
{We used \textit{Chandra} and XMM-\textit{Newton} archival data to derive the X-ray spectral shape and variability. 
We also obtained optical spectroscopy to detect the expected emission lines that are typically found  in Active Galactic Nuclei. And we finally used public ZTF imaging data to investigate the optical variability.}
{There is no evidence of multiple X-ray sources or extended component within CGCG 077-102 NED02. Single X-ray spectral
models fit well the source. Non-random 0.5-10 keV significant X-ray flux inter-observation X-ray variabilities were detected,
between $\sim$4~days for short term variations and up to $\sim$700~days for long term variations. Optical spectroscopy
points toward a passive galaxy for CGCG 077-102 NED01 and a Seyfert for CGCG 077-102 NED02. The classification of CGCG 077-102 NED02 is also consistent 
with its X-ray luminosity of over 10$^{42}$  erg s$^{-1}$. We did not detect short-term variability in the optical ZTF light curves. However, we found a significant long-term stochastic variability in the $g$-band that can be well described by the damped random walk model with a best-fitted characteristic damping timescale of $\tau_{\mathrm{DRW}}=30_{-12}^{+28}$ days. Finally, the CGCG 077-102
galaxy pair is deeply embedded within the Abell 2063 potential, with a long enough history within this massive structure to have underwent the cluster influence for a long time.}
{Our observations point toward a moderatly massive black hole in the center of CGCG 077-102 NED02, of $\sim 10^6$ M$\odot$. As compared to another similar pair in the literature, CGCG 077-102 NED02 is not heavily obscured, perhaps due to surrounding intra cluster medium ram pressure stripping.}


\maketitle

\section{Introduction}
\label{sec:intro}

One of the unsolved puzzles of galaxy formation and evolution is the link between the central supermassive black hole (SMBH hereafter)
in the galaxy core and the galaxy itself: \cite{Latif2022Natur,Wise2019Natur,2020MNRAS.499.2380S} and references therein. The existence of the SMBH can give rise to an Active Galactic Nuclei (AGN hereafter), the activity being enhanced by the fact it belongs to a pair of galaxies, see e.g. an extensive review of the literature as of 2019: \cite{2019ApJ...875..117P}. 

In this framework, a ``pair'', as defined by \cite{2021MNRAS.504..393G}, is made of two galaxies separated by less than 150\,kpc. \citeauthor{2021MNRAS.504..393G} searched for X-ray emission from 32 galaxy pairs that were serendipitously observed by \textit{XMM-Newton} with at least one member classified by the SDSS as an AGN. By using the X-ray emission to classify a galaxy as an AGN, they found extra AGNs. However,
 \cite{2019ApJ...875..117P} carried out a search similar to that of \cite{2021MNRAS.504..393G} and suggested that AGNs in post-merger systems are heavily obscured, so post-mergers can be missed in searches relying on \xmm\ detections. We note that, even though the \cite{2019ApJ...875..117P} search found a pair within one degree of the pair we discuss here (CGCG 077-102 hereafter), their search missed CGCG 077-102, because neither of its members was classified as an AGN in the SDSS database.

\begin{figure}[ht]
\centering
\includegraphics[width=8.8cm]{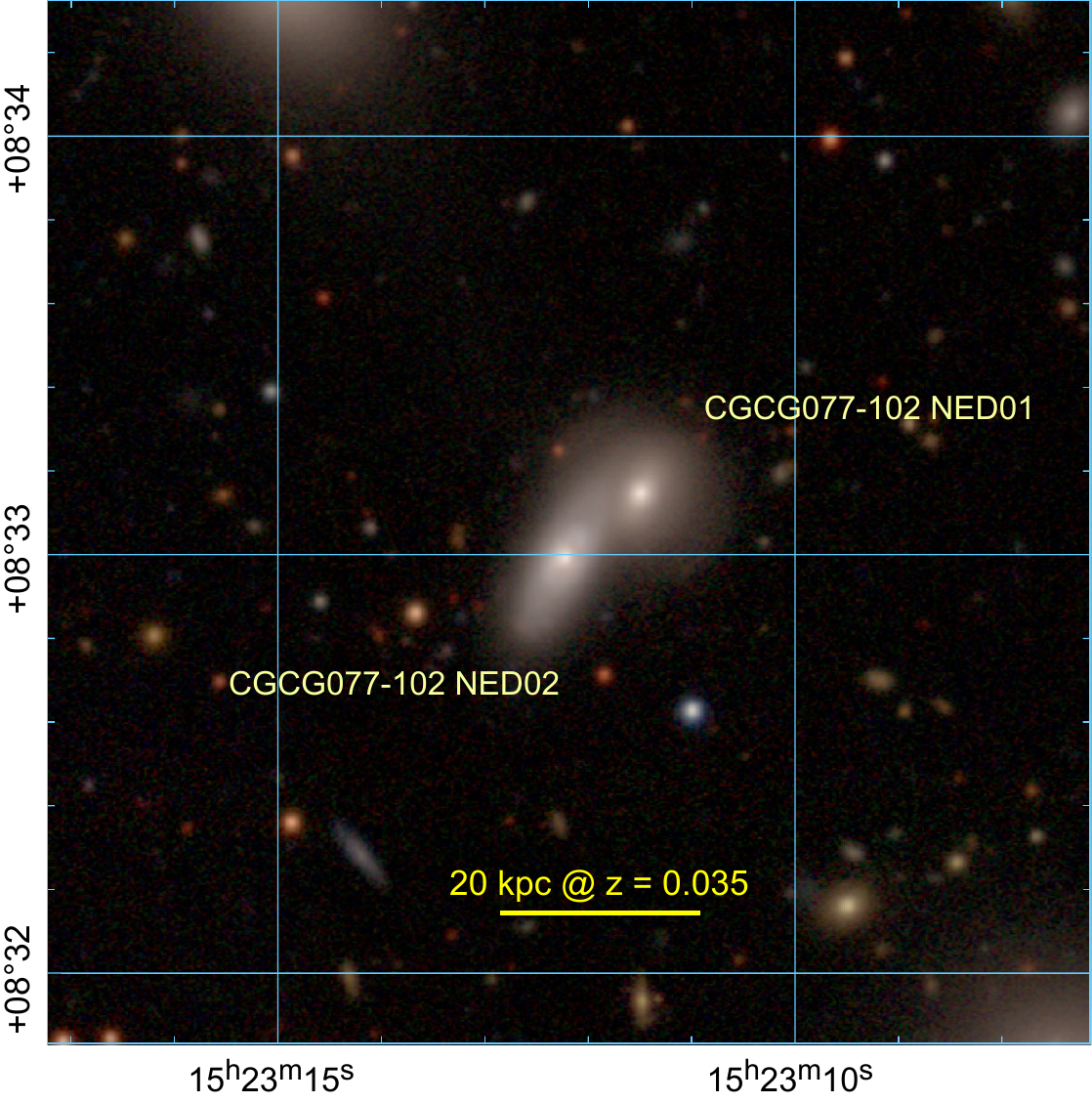}
\caption[]{DESI Legacy $gri$ image of the galaxy pair CGCG 077-102. The coordinates are equatorial J2000.0.}
\label{fig:map0}
\end{figure}

Our attention was brought to this pair because, while conducting a search for tidal disruption events (TDEs) in the archival \cha\ and \xmm\ databases \citep{2014MNRAS.444..866M,2013MNRAS.435.1904M,2010ApJ...722.1035M},
 we discovered a variable X-ray source. The X-ray source position was within 0.\!\arcsec2 of the nucleus of a galaxy located in
the rich cluster of galaxies Abell 2063 ($z = 0.035$, A2063 hereafter). This galaxy is a member of the galaxy pair 
CGCG 077-102. The 
two member-galaxies (CGCG 077-102 NED01 and CGCG 077-102 NED02: NED01 and NED02 hereafter) have a very similar appearance in the optical (see Fig.\ref{fig:map0}) and NED02 is classified as Sab and NED01 as S? in the Siena Galaxy Catalog \footnote{https://www.legacysurvey.org/sga/sga2020/}. 
In hindsight, the CGCG 077-102 pair could have been found in a search that included both a combination of a mid-IR and ``hard'' (2-10 \,keV) X-ray detection.  In this framework, a search in the WISE database (see Summary and Conclusions) shows that the AGN of the CGCG 077-102 pair (NED02) stands out from the non-AGN by being more than 10 times more luminous in the WISE channel 4 (about 22\,$\mu$).  Thus, the CGCG 077-102 pair discussed here can be considered typical for the pairs of \citeauthor{2021MNRAS.504..393G}
except that the separation in projection is smaller: only 10 \,kpc, and the AGN of the pair (see below) was not classified by the SDSS software because no spectrum was recorded in the SDSS database. Since the variability did not show a classic t$^{-5/3}$, (e.g. Gezari 2021) we discounted the assignment of a TDE to the origin of the X-ray flux from NED02.

The CGCG 077-102 pair is also similar in many ways to the  AGN-non-AGN pair J085953.33+131055.3 described in \cite{2019ApJ...875..117P}, see their Figs 2 and 3. For, except for J085953.33+131055.3 being in the field vs being in an X-ray luminous Abell cluster,  both appear to the eye to be spirals. The X-ray emitting member of the pair is the largest of the two, the redshifts are nearly identical (0.0308 for J085953.33+131055.3 vs $\sim$0.035 for CGCG 077-102) and the projected separations are both about 10\arcsec ($\sim$7kpc). A detailed study of the CGCG 077-102 pair can then be used to understand how being in a cluster can affect the accretion flow onto the central SMBH of the AGN member of the pair and the central obscuration.  

We were therefore motivated to make optical observations to determine if
only the X-ray active galaxy of the pair (NED02) also was the only one to exhibit emission lines
characteristic of AGN. In the case of this pair, as we show in this paper, the X-ray luminosity and the optical
emission line spectra are linked.

The outline of this paper is as follows. In section \ref{sec:xray}, we investigate the X-ray nature of NED02. In section \ref{sec:spec}, 
we detail the optical spectroscopy of the CGCG 077-102 pair. Section \ref{sec:dynamical} gives an analysis of the dynamical status of the pair 
within the A2063 cluster. Section \ref{sec:optical} shows the optical light curve analysis, and last section is the conclusion. Two appendices describe ancillary results.

All along that paper we use the following cosmological parameters: 
$H_0 = 69.6\,$km~s$^{-1}$~Mpc$^{-1}$,  $\Omega _\Lambda =0.714$ and 
$\Omega _m=0.286$.

\section{The X-ray nature of CGCG 077-102 NED02}
\label{sec:xray}
\begin{figure}[ht]
\centering
\includegraphics[width=8.0cm]{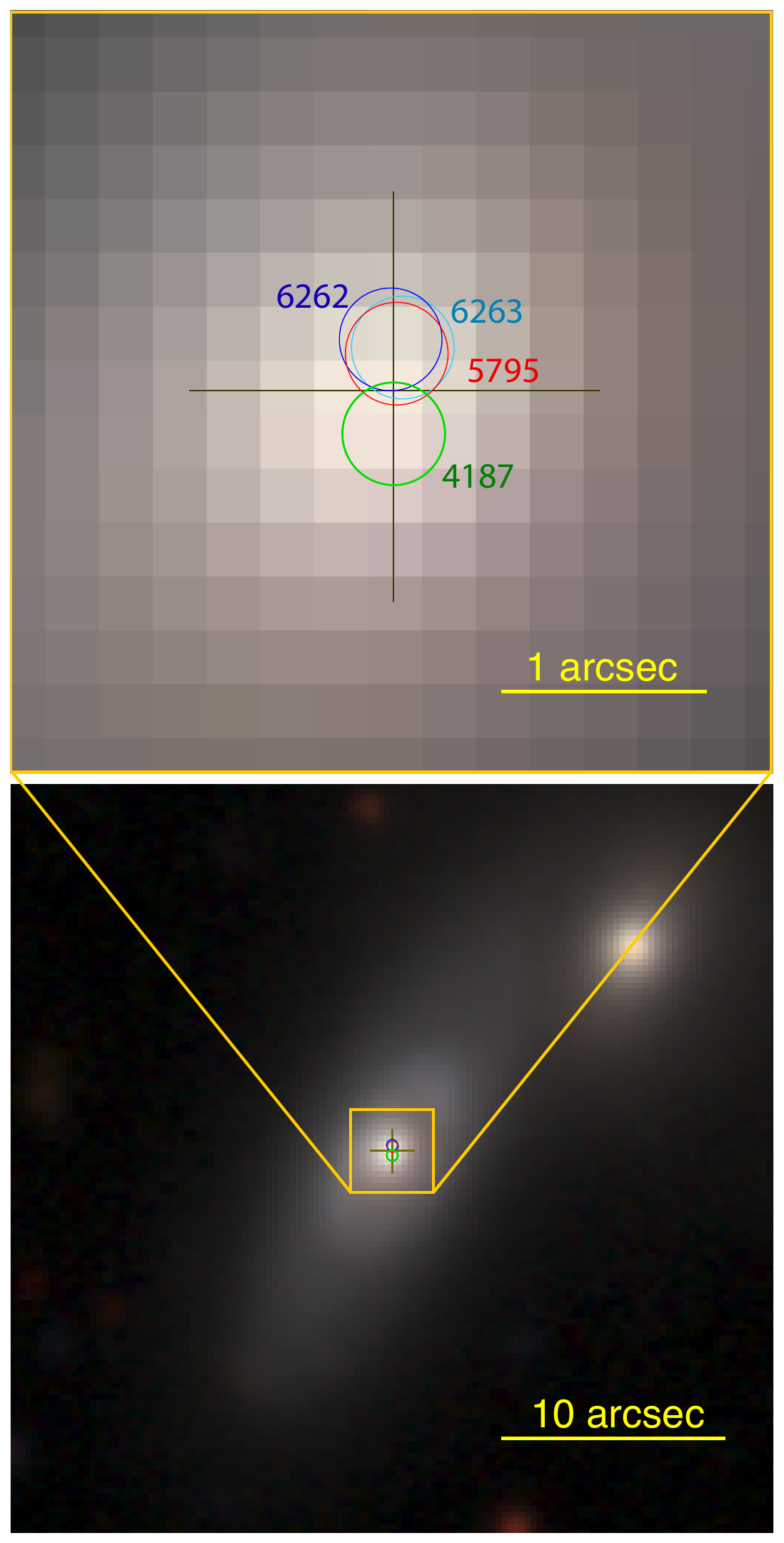}
\caption[]{The center of the X-ray source in each \textit{Chandra} pointing (green, red, blue and cyan circles of radius 0.25\arcsec) superposed on a true-color optical \textit{gri} DESI Legacy Survey image of CGCG 077-102 galaxy pair. Below: NED01/NED02 field. Above: zoom-in on NED02. The cross represents the optical center of the galaxy. The pixel size is 0.262\arcsec~ and North is up and East is left on both panels.}
\label{fig:centerSrcsX}
\end{figure}

NED02 has an X-ray source that is present in both 4XMM-DR11 Catalogue \citep{2020A&A...641A.136W} and \textit{Chandra} Source Catalog 2.0 \citep{2020AAS...23515405E} (see Fig.~\ref{fig:centerSrcsX}). The XMM-\textit{Newton} catalogue has 3 measures and the CSC-2 has 4 pointings that we summarize in Table~\ref{tab:catalogX}.

\begin{table*}[ht]
\caption[]{Data from XMM-\textit{Newton} 4XMM-DR11 and \textit{Chandra} CSC-2 catalogues. The flux is in units of $10^{-12}$erg~s$^{-1}$~cm${^{-2}}$; for XMM-\textit{Newton}, in the [0.2--12 keV] band, while the \textit{Chandra} flux is in the [0.5--7.0 keV] band. The off-axis distance, off\_ax, is in arcmin and the net exposure time, exp, is given in ks.}\label{tab:catalogX}
\centering
\begin{tabular}{ccccc| cccccc}
\hline
\hline
\multicolumn{5}{c|}{4XMM-DR11} & \multicolumn{6}{c}{CSC-2} \\
\hline
ObsID & Date obs. & exp & off\_ax & flux$_{\rm XMM}$ & ObsID & Date obs. & exp & off\_ax & Detector& flux$_{\rm CSC}$ \\
\hline
0200120401 & 2005-02-17 &  17.4 & 3.76 & 2.229 & 4187 & 2003-04-20 & 8.78 & 4.834 & ACIS-I3 & 0.679 \\
0550360101 & 2008-07-23 &  24.4 & 2.95 & 1.517 & 5795 & 2005-04-02 & 9.91 & 5.144 & ACIS-S3 & 1.222 \\
0782531001 & 2017-01-25 &  12.0 & 11.7 & 2.054 & 6262 & 2005-03-28 & 14.2 & 5.143 & ACIS-S3 & 1.329 \\
           &            &       &      &       & 6263 & 2005-03-29 & 16.8 & 5.143 & ACIS-S3 & 1.393 \\
\hline
\hline
\end{tabular}
\end{table*}

In both catalogs, the X-ray emission associated with the galaxy NED02 is classified as a point-source. In order to check the automatic pipeline results, we have downloaded the publicly available \textit{Chandra} observations and reprocessed the data. We have chosen to use the \textit{Chandra} data because of the better spatial resolution and the total exposure time (adding all observations) is comparable to the XMM total exposure. We have, thus, combined all four observations listed in Table~\ref{tab:catalogX} (the PI was C. Sarazin in all 4 observations).

The earlier observation from 2003 was done with ACIS-I in very faint mode, with the target on the CCD number I3. The other 3 observations in 2005 were done with ACIS-S, also in very faint mode, using the back-illuminated CCD S3. Since the aimpoint was the center of Abell 2063, our object of interest, NED02 was always significantly off-center (see Table~\ref{tab:catalogX}).

For the reprocessing of the downloaded \textit{Chandra} data, we have followed the standard procedure described in the CXC Science Threads\footnote{\texttt{https://cxc.harvard.edu/ciao/threads}}, using CIAO~4.12 and CALDB~4.9.3.
We checked for high background periods (flares) in the [0.5--7.0 keV] band, by producing light-curves with 100s bin and masking the bright point-sources. No such period of flares was found in any of the four exposures.

\begin{figure}[ht]
\includegraphics[width=8.6cm]{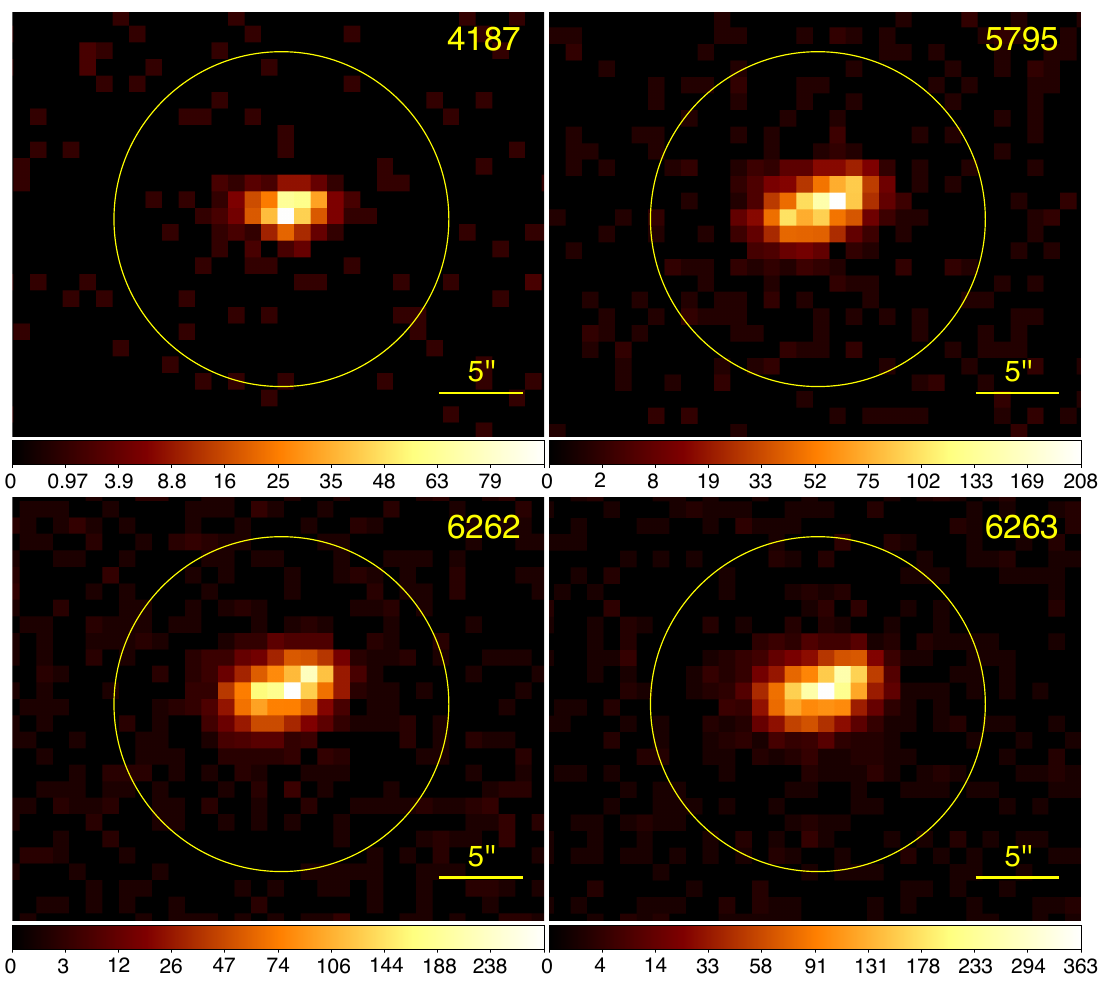}
\caption[]{Exposure-map corrected image of each \textit{Chandra} observations. They are all broad-band, [0.5--7.0 keV]. The color scale is \texttt{sqrt} and color bar is in counts per pixel (1 pixel = 0.984\arcsec; for this image the ``natural'' ACIS pixel size was binned by a factor 2). The scale is in counts per pixel. North is up and East is left on all panels.}
\label{fig:XrayImgs}
\end{figure}

We then produced a broad-band, exposure-map corrected image for each exposure (see Fig.~\ref{fig:XrayImgs}). The X-ray source in exposure \#4187 has a different position angle with respect to the other three pointings, that is why it seems rotated compared to the other images (all images have the same orientation and scale in Fig.~\ref{fig:XrayImgs}).

We now expand on the potential multiplicity of the X-ray source, as it is crucial to study a potential time-variability.

\subsection{Single or multiple X-ray sources within CGCG 077-102 NED02?}

We have used \texttt{ChaRT} v2 to simulate the PSF at the position of the X-ray source and MARX~5.5.1 for creating a simulated event file of a single point-source. The point source is modeled using the spectrum of NED02 for each exposure and an observation is simulated using the same characteristics of the real observations (offaxis distance, roll-angle, exposure time, very faint mode, and celestial coordinates). 

\begin{figure}[ht]
\includegraphics[width=8.6cm]{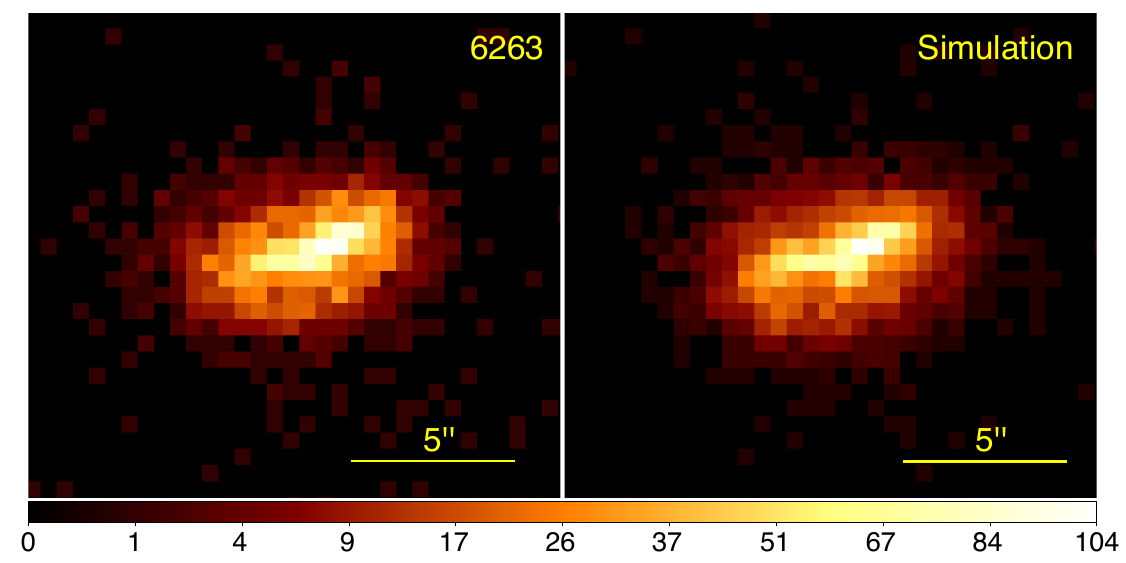}
\caption[]{Image of the \textit{Chandra} observation \#6263 (left) compared to the ChaRT+MARX simulation (right). Both images are in the [0.5-7.0 keV] band with plate scale of 0.492\arcsec, the unbinned ACIS pixel size. North is up, East is to the left.}
\label{fig:XrayPSF}
\end{figure}

In Figure~\ref{fig:XrayPSF} we show an example of a simulated and an observed image corresponding to observation \#6263; for the other exposures, the results were very similar. Clearly, the X-ray source can be easily accounted by a single spatially unresolved source, affected by \textit{Chandra} high resolution mirror assembly (HRMA) telescope point spread function (PSF). There is no evidence of multiple X-ray sources or an extended component.
However, an on-axis Chandra observation would be the ideal way to use imaging to confirm our supposition that there is a single 
X-ray source located at the core of NED02.

\subsection{X-ray Spectral analysis}

Next, we have checked whether a single absorbed power-law emission could describe the X-ray source. We have extracted the spectrum, for each exposure, inside a circle of radius 7" centered at 15h23m12.21s, +08$^\circ$32$^\prime$59.7" (J2000). The background was also extracted from each exposure, within two apertures of radius 16.4" centered at 15h23m14.82s, +08$^\circ$33$^\prime$19.4" and 15h23m10.06s, +08$^\circ$32$^\prime$35.1". However, in the case of exposure \#4187, accounting for its mismatched position angle, we have taken a source region of radius 7" at 15h23m12.23s, +08$^\circ$32$^\prime$59.6" and two background apertures of radius 16.4" at 15h23m14.47s, +08$^\circ$33$^\prime$21.2" and 15h23m09.72s, +08$^\circ$32$^\prime$37.0".

For a source with a flux 
$\sim$$10^{-12}\,$erg~s$^{-1}$~cm$^{-2}$, some pileup is to be expected. Using the task \texttt{pileup\_map} we verified that the level, for the ACIS-I observation \#4187, is about 7\% on the brightest pixel, dropping to less than 1\% four pixels away from the brightest one. For the ACIS-S observations, the brightest pixel have a pileup around 13\%, dropping to 2\% four pixels away from the brightest. The relatively low pileup is because the source is far from the \cha\ ``aim point,'' where the effective area is lower. Therefore, such a minimal pileup may be ignored and did not affect our \cha\ spectroscopic analysis nor our interpretation of the results.

For each \textit{Chandra} observation, we computed the rmf and arf files using the CIAO script \texttt{specextract}. The spectral fits were done using XSPEC 12.13.0c, with the spectra channels grouped with a minimum of 15 counts per bin, but limited to the range [0.3--7.0 keV]. We have used for the statistical weight the ``Churazov'' method (using the mean value from adjacent energy channels, as recommended in the XSPEC manual), appropriate for small number statistics. The total net count (background subtracted) is shown in Table.~\ref{tab:fitXspec}.

\begin{figure}
\includegraphics[width=8.7cm]{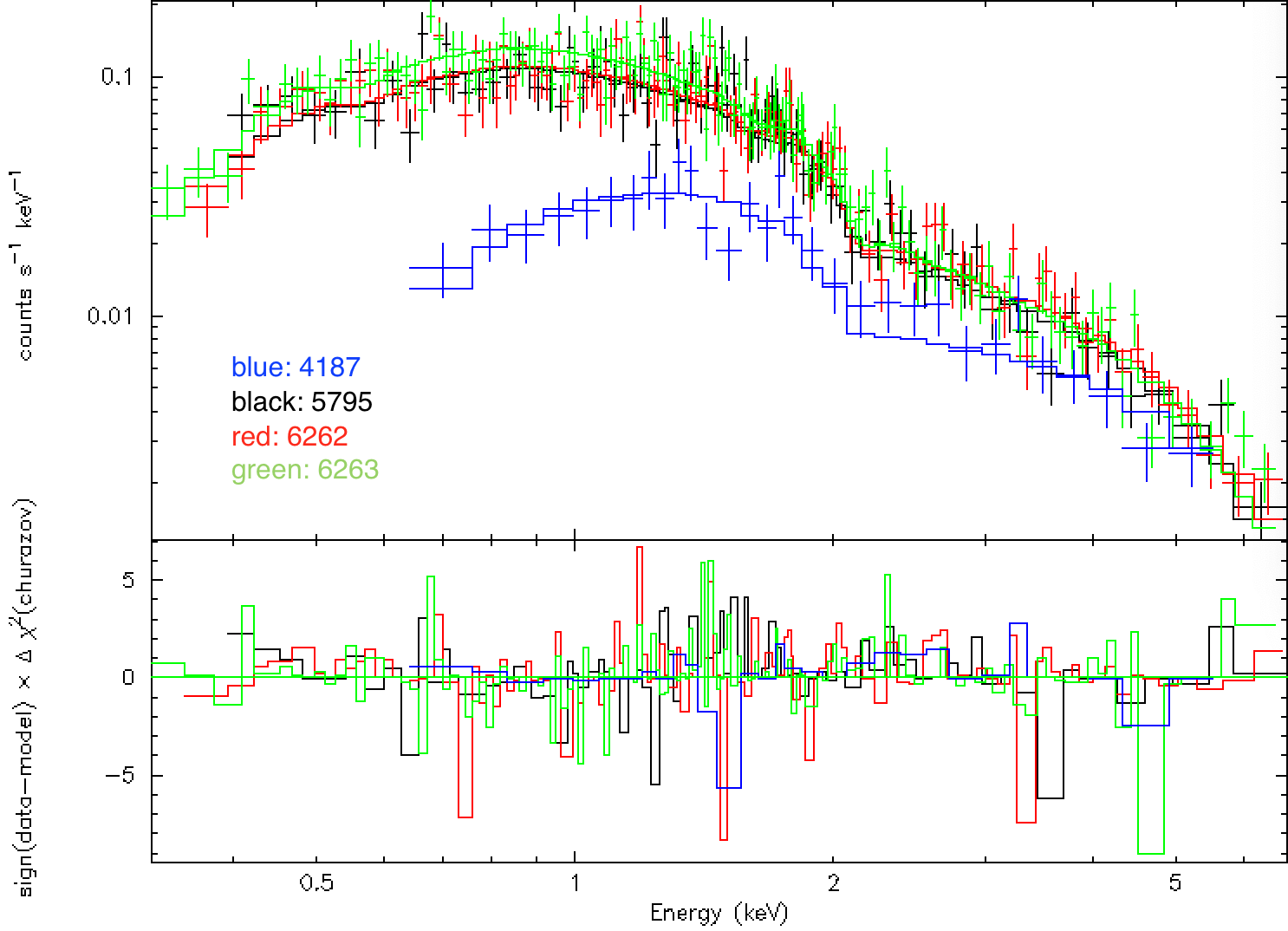}
\caption[]{\textit{Chandra} spectra analysis. \textsf{Top:} data with best-fit single power-law model. Each color line represents a different observation (identified by its ObsID, see Table~\ref{tab:catalogX}). Note that the ObsID 4187 spectrum (blue curve) was done with ACIS-I, while the other 3 with ACIS-S. \textsf{Bottom:} Residues (data $-$ model) in terms of the contribution to the $\chi^2$ for each energy bin in the spectrum.} 
\label{fig:XraySpectraXspec}
\end{figure}

We used an absorbed (Galactic and intrinsic) power-law model (xspec \texttt{pow}) to fit the observed spectra. The Galactic hydrogen column density was fixed using the HI4PI \citep{2016A&A...594A.116H}, $N_{H} = 2.61 \times 10^{20}$~cm$^{-2}$. The resulting best fit is given in Table~\ref{tab:fitXspec} and shown in Fig.~\ref{fig:XraySpectraXspec}. We note that two-component fittings (i.e., fitting with two spectral components, for instance, adding two power-laws) do not give significantly better results and we will assume for now that we are dealing with a single component X-ray source.

We note that the best-fit free parameters from the ObsID \#4187 are slightly different from the values obtained by fitting the other 3 exposures (although they are all well within 1-$\sigma$ confidence interval). ObsID \#4187 was done with ACIS-I in faint mode, while the other 3 were done with ACIS-S in very faint mode. Moreover, ObsID \#4187 is shallower than the other observations, what explains the larger error bars. 

\begin{table*}[ht]
\centering
\caption[]{Results of the single power-law spectral fit for observations \#4187, 5795, 6262, and 6263. $\Gamma$ denotes the photon index of the power-law and $N_H$ gives the intrinsic hydrogen column density of the source. Fluxes derived from the best fits are in units of $10^{-12}$ \ecmss. Luminosities are in $10^{42}\,$erg~s$^{-1}$ and ``cnt'' is the net count (background subtracted)}. Error bars are 90\% confidence level.
\label{tab:fitXspec}
\begin{tabular}{ccccccccc}
\hline
ObsID&$\Gamma$ & $N_{H}$ & $\chi^2$/d.o.f & flux & flux & luminosity & luminosity & cnt\\
     &         & [$10^{20}$cm$^{-2}$]&   & [0.5--7.0 keV] & [0.2--12.0 keV] & [0.5--7.0 keV] & [0.2--12.0 keV] & \\
\hline
\noalign{\vskip 0.1cm} 
4187  & $1.503^{+0.268}_{-0.225}$& $7.807^{+15.008}_{-7.807}$ & 24.11/28   & $0.701^{+0.066}_{-0.087}$ & $0.983^{+0.147}_{-0.155}$ & $1.96^{+0.18}_{-0.24}$ & $2.75^{+0.41}_{-0.43}$ & ~526 \\[5pt]
5795  & $1.748^{+0.137}_{-0.091}$& $0.924^{+3.964}_{-0.924}$  & 93.07/82   & $1.125^{+0.053}_{-0.076}$ & $1.510^{+0.105}_{-0.134}$ & $3.15^{+0.15}_{-0.21}$ & $4.23^{+0.29}_{-0.38}$ & 1700 \\[5pt]
6262  & $1.679^{+0.027}_{-0.063}$& $0.255^{+2.719}_{-0.255}$  & 123.28/119 & $1.213^{+0.051}_{-0.059}$ & $1.651^{+0.097}_{-0.097}$ & $3.40^{+0.14}_{-0.17}$ & $4.62^{+0.27}_{-0.27}$ & 2541 \\[5pt]
6263  & $1.875^{+0.095}_{-0.091}$& $2.332^{+2.394}_{-2.246}$  & 150.15/141 & $1.249^{+0.050}_{-0.050}$ & $1.635^{+0.091}_{-0.089}$ & $3.50^{+0.14}_{-0.14}$ & $4.58^{+0.25}_{-0.25}$ & 3334 \\[5pt]
\hline
\end{tabular}
\end{table*}

\subsection{X-ray Variability}

\begin{figure}[ht]
\centering
\includegraphics[width=9cm]{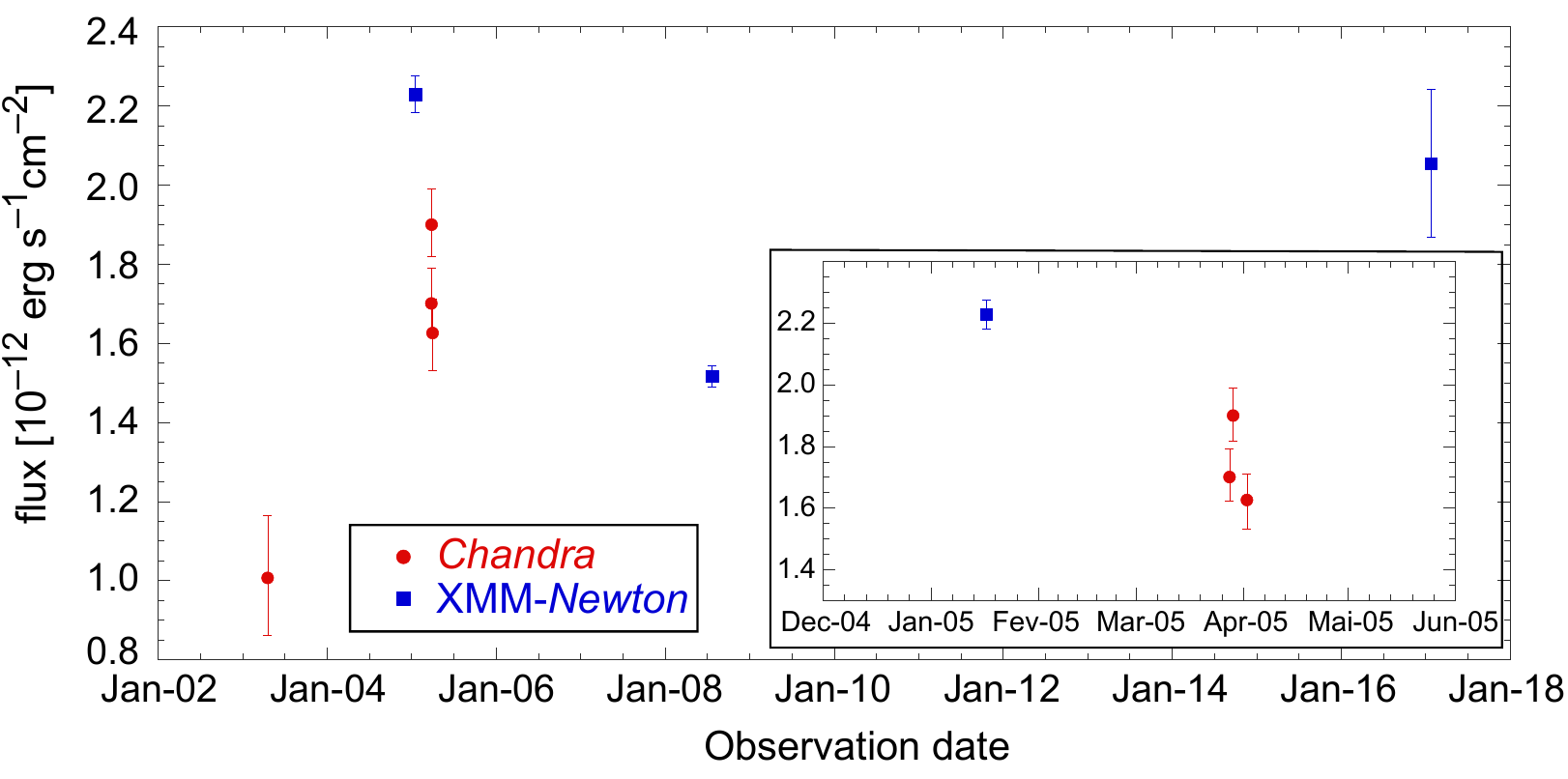}
\caption[]{X-ray flux [0.2--12.0 keV], extrapolated from the fitted spectrum for Chandra data and from the archive for XMM-\textit{Newton} data, as a function of time. The inset at the bottom right zooms on the period from December 2005 to June 2005.}
\label{fig:variab_Chandra_XMM_A2063}
\end{figure}

We initially made use of the XCTHULHU automated procedure for discovering tidal flare candidates in archival \cha ~
data\footnote{https://github.com/xcthulhu/Galaxy-Clusters}. Figure~\ref{fig:variab_Chandra_XMM_A2063} shows the X-ray flux from both \textit{Chandra} and XMM-\textit{Newton} observation as a function of time. 

To quantify the short-term X-ray variability in each \textit{Chandra} exposure, we have implemented the Gregory-Loredo algorithm via CIAO's \texttt{glvary} method.  An advantage of this implementation is that dither corrections for an exposure are applied via re-normalizing the light curve, so the variability is not flagged by any dither modulation across the ACIS chips.
In assigning a single variability index (in the range 0 to 10) to an observation, the algorithm conditions over:
\begin{itemize}
\item The probability that the source flux is not constant over the observation exposure time
\item The combinatoric ``total odds ratio" from binning $N$ events into $M$ time bins
\item The fractions of the light curve within 3$\sigma$ and 5$\sigma$, respectively, of the average count rate in a given exposure
\end{itemize}
An index $\ge6$ signifies that a given light curve is ``definitely variable" -- corresponding to high probability of non-constant flux ($P\sim1$) and/or high total odds ratio $O\gtrapprox2$. The $3\sigma/5\sigma$ light curve fractions are only relevant at low variability index.

For NED02, ObsIDs 4187, 6262, and 6263 were flagged as definitely not variable with index 0. However, ObsID 5795 was flagged as definitely variable with index 7 -- attributable to a non-negligible decrease in flux over its exposure time. As a control, we have tested the variability of the diffuse emission within the background apertures for the ObsID 6263 exposure, which \texttt{glvary} flagged as non-variable with index 0.

To assess the X-ray variability of NED02 in greater detail, we consider (1) the long-term variability across all 4 exposures and (2) the long-term variability across the most recent three exposures (omitting ObsID 4187).

For each case, the X-ray light curve was grouped into 500 s bins, and bins with zero count rate were discarded to consider only non-trivial X-ray detections in the variability analysis. To quantify the variability of this light curve, we have implemented the long-term variability measures described in \cite{2021A&A...650A.167D}. The results are shown in Table.~\ref{tab:var_ned02}.

\begin{table*}[ht]
\centering
\caption[]{Variability measures for the long-term NED02 X-ray light curves for the \textit{Chandra} exposures, quantifying the variability size and timing. \textbf{(a)} Reduced $\chi^2=\frac{1}{n-1}\sum_i^n\left(\frac{x_i-\overline{x}}{\sigma_i}\right)^2$ about the error-weighted mean count rate, and the corresponding $p$-value. $p$-values $<0.0027$ are deemed significant (i.e. non-variability is improbable) at a $3\sigma$ level. \textbf{(b)} Maximum error-normalized count rate change, in counts per second. \textbf{(c)} Ratio of maximum count rate to the minimum count rate (MR) and the propagated error (MRE). \textbf{(d)} Time interval for the largest error-normalized count rate change, in days. \textbf{(e)} Error-weighted mean count rate, in counts per second. \textbf{(f)} Maximum upward/downward count rate ratio $\left(\frac{x_i-\sigma_i}{x_j+\sigma_j}\right)_{max}$ for $i>j$ (EMFU), $i<j$ (EMFD). \textbf{(g)} Corresponding time intervals for EMFU and EMFD, in days. \textbf{(h)} Shortest time interval (in days) for a count rate increase (ET2U) or decrease (ET2D) by a factor $>2$, i.e. where $\frac{x_i-\sigma_i}{x_j+\sigma_j}>2$. If no such interval exists, we report as n.a. \textbf{(i)} Shortest time interval (in days) for a count rate increase (ET10U) or decrease (ET10D) by a factor $>10$, i.e. where $\frac{x_i-\sigma_i}{x_j+\sigma_j}>10$. \textbf{(j)} $Z$ test statistic for the Wald-Wolfowitz runs test. At 5\% significance ($|Z|<1.96$), one should accept a null hypothesis of random variability.}

\begin{tabular}{cccccccc}
\hline
       & \begin{tabular}[c]{@{}c@{}}$\chi^2_\nu$/\\$p$-value$^{(a)}$ \end{tabular} & \begin{tabular}[c]{@{}c@{}}MDDE$^{(b)}$\\ \end{tabular} & \begin{tabular}[c]{@{}c@{}}MR/\\MRE$^{(c)}$ \end{tabular} & \begin{tabular}[c]{c@{}}TMDDE$^{(d)}$\\ \end{tabular} & \begin{tabular}[c]{@{}c@{}}WMEAN$^{(e)}$\\ \end{tabular} & \begin{tabular}[c]{@{}c@{}}EMFU/\\EMFD$^{(f)}$ \end{tabular} & \begin{tabular}[c]{@{}c@{}}TEMFU/\\TEMFD$^{(g)}$ \end{tabular}  \\ \hline
All 4 ObsIDs & $15.662/$ & 9.129 & $11.400/$ & 708.51 & 0.132 & 5.588/ & 709.49/\\
 & $\sim$0 &  & 14.021 &  &  & 2.033 & 4.24\\
4187 Omitted & $2.420/$ & 5.515 & 2.553/ & 4.24 & 0.183 & 1.766/ & 0.79/\\
 & $5.327\times10^{-12}$ &  & 0.070 &  &  & 2.033 & 4.24\\
 \hline
\end{tabular}
\begin{tabular}{cccc}
       & \begin{tabular}[c]{@{}c@{}}ET2U/\\ET2D$^{(h)}$ \end{tabular} & \begin{tabular}[c]{@{}c@{}}ET10U/\\ET10D$^{(i)}$ \end{tabular} & \begin{tabular}[c]{@{}c@{}}$Z_{WW}$$^{(j)}$ \end{tabular}  \\ \hline
All 4 ObsIDs & 708.49/ & n.a./ & -10.768\\
  & 4.24 & n.a. & \\
4187 Omitted & n.a/ & n.a/ & -12.658\\
 & 4.24 & n.a & \\
 \hline
\\
\end{tabular}
\label{tab:var_ned02}
\end{table*}

We therefore conclude that we have a significant inter-observation X-ray variability that is likely non-random.
We will now investigate the variable nature of NED02 in the optical wavelengths.

\section{Optical spectroscopy of the CGCG 077-102 galaxy pair}
\label{sec:spec}

NED01 galaxy member of the CGCG 077-102 pair is a $z \sim 0.03439$ passive galaxy (star formation rate basically
equal to 0 from SDSS DR17 database: StarformingPort model
\footnote{The StarformingPort and PassivePort models from the SDSS database are
estimating stellar masses of galaxy for SDSS and BOSS galaxies with the
Portsmouth method and star-forming and passive models. Stellar masses come
from the method of Maraston \cite[]{2009MNRAS.394L.107M}. These fit stellar
evolution models to the SDSS photometry, using the known spectroscopic
redshifts and assuming the Kroupa initial mass function \cite[]{2001MNRAS.322..231K}.
For star-forming case, the star-formation model uses a metallicity
and one of three star-formation histories: constant, truncated, and
exponentially declining. 
For passive models, the model is an instantaneous burst stellar population
whose age is fit for (with a minimum allowed age of 3 Gyrs). The population is
97$\%$ solar metallicity and 3$\%$ metal-poor, by mass. 
We refer the reader to https://skyserver.sdss.org/dr18/MoreTools/browser\#} for
more details. ), observed by the SDSS survey (obj. id: 1937715310949853184).
Its population age is estimated to 9.75 (between 8.5 and 10.75) Gyrs (PassivePort model).


NED02, the other pair member, is at z$\sim$0.03662 (e.g. Lee et al. 2017). To the best of our knowledge, there is
no publicly available optical spectrum for this object, even in the SDSS database.

In order to deeply investigate the spectral characteristics of the CGCG 077-102 pair, we mapped 
it at the Observatoire de Haute Provence with two instruments.

\subsection{MISTRAL observations}

\begin{figure*}[ht]
\includegraphics[width=18.cm,angle=0]{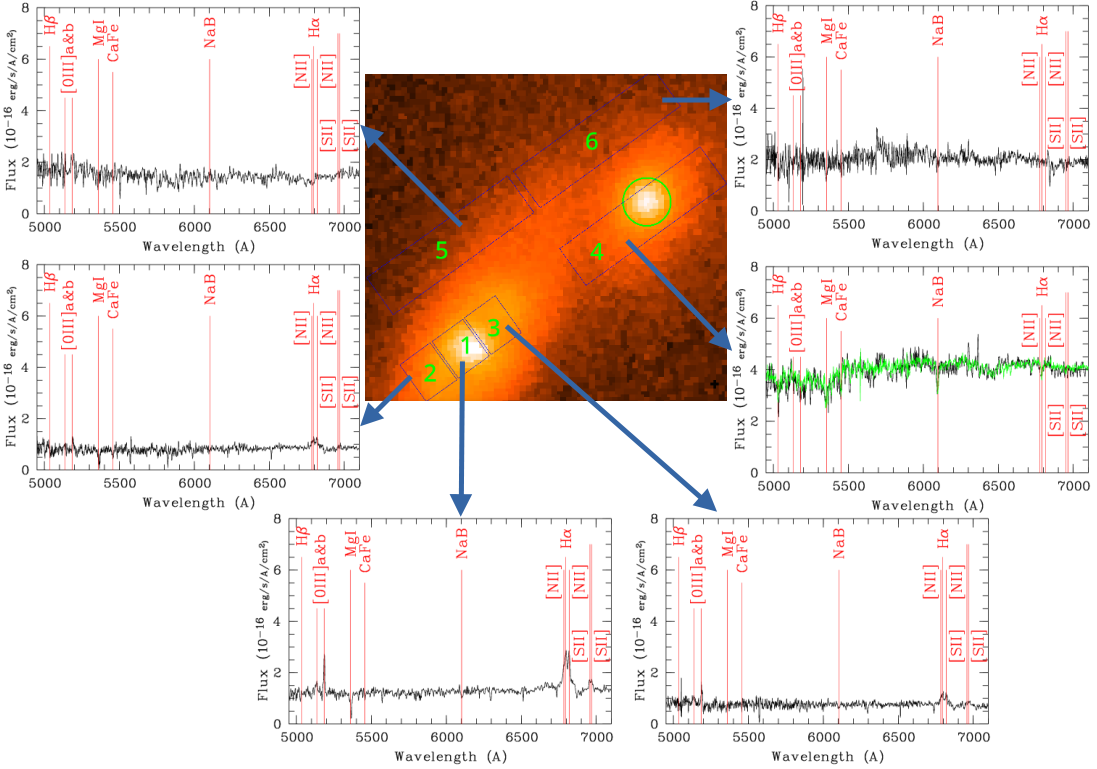}
\caption{\label{fig:n4} MISTRAL spectroscopic observations of six different regions of CGCG 077-102. Spectra are flux calibrated.
Location of the SDSS fiber is shown as the green circle. SDSS spectrum (green spectrum) is superimposed to the MISTRAL spectrum (black spectrum) 
for region \#4.}
\end{figure*}

MISTRAL is a low resolution single-slit spectro-imager\footnote{http://www.obs-hp.fr/guide/mistral/MISTRAL$\_$spectrograph$\_$camera.shtml)}. We used it in February 2022 to observe the CGCG 077-102 pair. 
We selected the blue setting
(1 hour exposure, R$\sim$750, covered wavelength range: [4200, 8000]A) to observe two different slits (see Fig.~\ref{fig:n4}). The MISTRAL slit aperture is fixed to 1.9\arcsec. Spectra were wavelength-calibrated using a combination of Hg, Ar, and Xe lamps. Wavelength calibration exposures were taken just after the science exposures, before making any significant telescope and instrument move. Flux calibration was based on the simultaneous
observation of the HD 289002 spectrophotometric standard.

Data reduction was performed with the MISTRAL dedicated night spectral python-based data reduction tool. This code is based on the Automated SpectroPhotometric Image REDuction package \cite[ASPIRED\footnote{\url{https://github.com/cylammarco/ASPIRED}}]{2022ASPC..532..537L} and has been tuned to the MISTRAL needs.

The first observed slit was placed to cover the central area of the two galaxies. The second one was to sample the northern outskirts
of the pair. We obtained good enough signal-to-noise to extract spectra in six different regions (see Fig.~\ref{fig:n4} and Tab.~\ref{tab:obsspectro}). 

First, region \#4, covering the center of NED01 (1.9\arcsec$\times$11\arcsec) shows a MISTRAL spectrum very similar to the SDSS spectrum (see Fig.~\ref{fig:n4}) 
when rescaled to take into account the different apertures (SDSS is a 3\arcsec circle, different from the MISTRAL extraction area).
No emission lines are visible, and we detect deep absorption lines (H$\beta$, MgI, CaFe, NaD, and perhaps even H$\alpha$).
Region \#6 does not show either any emission line. Absorption lines as H$\beta$, MgI, and NaD are visible.
We therefore confirmed the passive status of NED01. None of the explored regions in this galaxy show any sign of gas ionization.

The situation is clearly different for NED02. The external area (region \#5: 1.9\arcsec$\times$11\arcsec)
does not show any emission lines. We may detect MgI and Ca/Fe. The three other regions (\#1, 2, 3: 1.9\arcsec$\times$ 2.4\arcsec) exhibit clear [OIII],
H$\alpha$, and [SII] emissions. More remarkably, regions \#1 and 3 exhibit prominent [NII]@6585A emissions, as strong as the 
H$\alpha$ line itself. We detect absorption MgI, Ca/Fe and NaD lines in regions \#1 and 3.
We also note that H$\beta$ is not present, nor in emission or absorption in regions \#1, 2 and 3.
It is therefore clear that emission lines are only present close to the NED02 center (see also below for high resolution integral field GHASP spectroscopic observations). 
We note that region \#1 includes the X-ray source detected in \cha ~ data.

\begin{table}[ht]
\caption[]{Integrated and surface brightness magnitudes of the 6 spectroscopically-sampled regions (see Fig.~\ref{fig:n4}). We also give the angular size of the regions. Magnitudes are computed from the public SDSS g'-band images of the region.}\label{tab:obsspectro}
\centering
\begin{tabular}{cccccc}
\hline
\hline
Region id & g' integrated & g' surface brightness & size\\
          & mag & mag/arcsec$^2$ & arcsec\\
\hline
1 & 17.4	& 19.5	& 1.9$\times$2.4 \\
2 & 18.6	& 20.7	& 1.9$\times$2.4 \\
3 & 18.1	& 20.2	& 1.9$\times$2.4 \\
4 & 17.0	& 20.7	& 1.9$\times$11 \\
5 & 18.9	& 22.6	& 1.9$\times$11\\
6 & 18.7	& 22.4	& 1.9$\times$11\\
\hline
\hline
\end{tabular}
\end{table}

\begin{table}[ht]
\caption[]{H$\beta$, [OIII]@5007, H$\alpha$, and [NII]@6584 fluxes measured in region \#1, in units of erg/s/cm$^2$.}\label{tab:fluxspectro}
\centering
\begin{tabular}{cccc}
\hline
\hline
 H$\beta$ & [OIII]@5007 & H$\alpha$ & [NII]@6584 \\
\hline
 $\leq$ 3.$\times$7.7 10$^{-17}$	& 1.5 10$^{-15}$ &  4.1 10$^{-15}$ & 5.0 10$^{-15}$ \\
\hline
\hline
\end{tabular}
\end{table}

These emission lines coming preferentially from the NED02 center are possibly due to an AGN in the galaxy center. However,
the galaxy pair probably being interacting, we may have some merging-induced star formation in NED02. To check this hypothesis,
we first used the AMAZED spectral fitting code 
\cite[e.g.,][]{2019ASPC..521..398S}.
Among the 21 available in the code, the template offering the best 
fit is a QSO, with a metalicity of 0.5 and a stellar population age of 5 Gyrs.

\begin{figure}[ht]
\includegraphics[width=9.cm,angle=0]{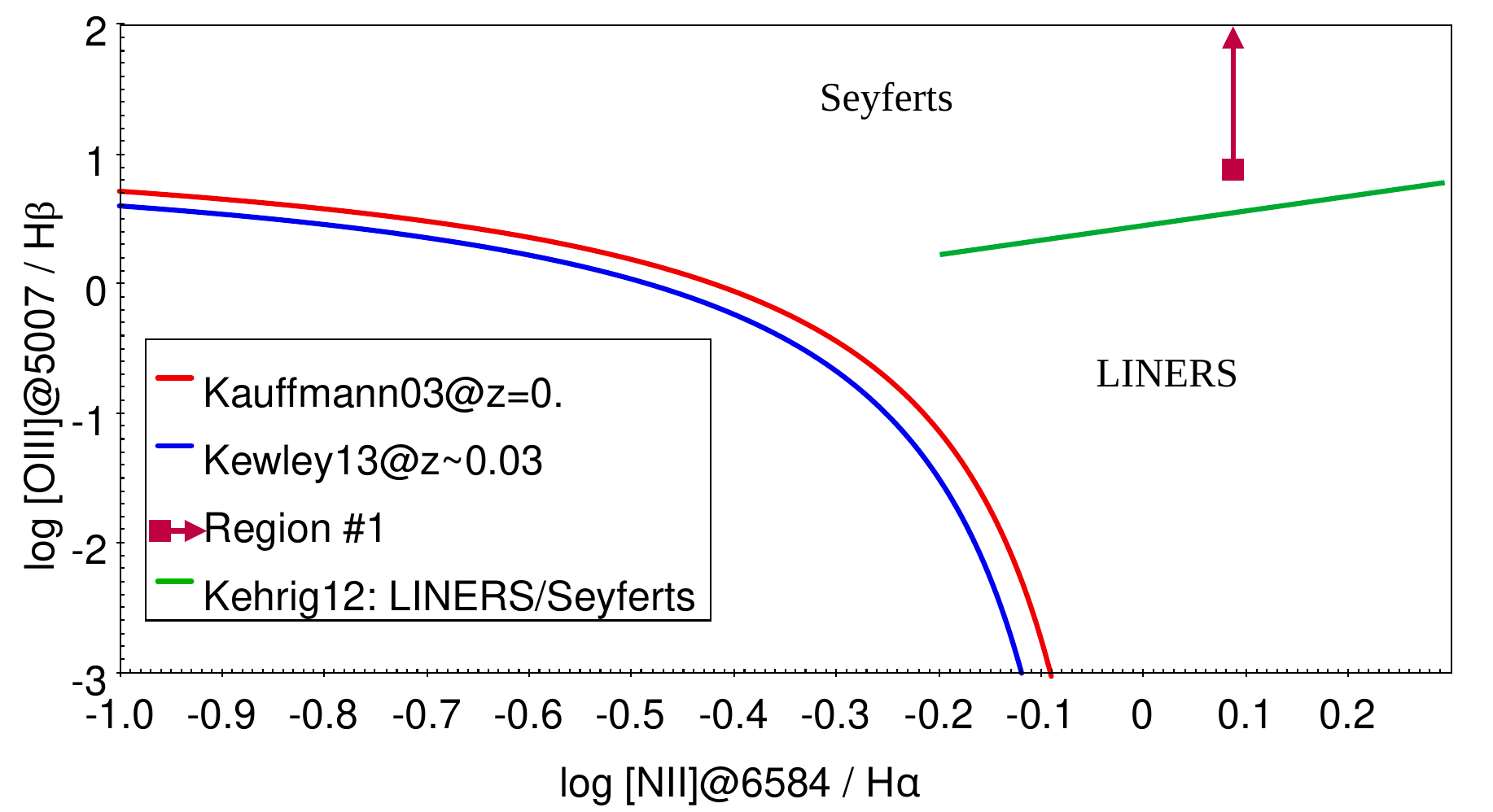}
\caption{\label{fig:n5} log([OIII]5007/H$\beta$) versus log([NII]6584A/H$\alpha$) for MISTRAL region \#1 spectrum. From 
\cite{2003MNRAS.346.1055K}
and 
\cite{2013ApJ...774..100K}
areas under the red and blue curves are for normal galaxies, while areas above these curves are for active objects. The green line shows the delimitation between Seyfert and LINERS galaxies 
\citep{2012A&A...540A..11K}.
The vertical dark magenta arrow is to indicate that
NED02 Region 1 has detectable H$\alpha$, N[II]6584 and O[III]5007 emissions but no detectable H$\beta$ flux. The bottom of this line shows the minimal value from the lower detection limit of H$\beta$.}
\end{figure}

We then measured the fluxes under the lines (see Table \ref{tab:fluxspectro}) using the SLINEFIT 
\cite[e.g.][]{2018A&A...618A..85S}
public code\footnote{https://github.com/cschreib/slinefit/blob/master/README.md} to generate line ratios. 
We used the H$\alpha$ / NII@6584A and O[III]@5007A / H$\beta$ lines ratio in order to place NED02 (region \#1) in the Fig.~\ref{fig:n5} diagram. H$\beta$ being 
not detected, we used instead the 3$\sigma$ noise within the [5000;5100]A range in order to generate a lower limit for O[III]@5007A / H$\beta$. Fig.~\ref{fig:n5} also shows the 
\cite{2003MNRAS.346.1055K}
and 
\cite{2013ApJ...774..100K}
separations between normal galaxies and active objects. 

We see that the NED02 
region \#1 has optical AGN characteristics. From 
\citeauthor{2014MNRAS.437.2376R} (2014: see their Fig 1),
\citeauthor{2017FrASS...4...34M} (2017: see their Fig 1), and 
\cite{2012A&A...540A..11K}
, we classify NED02 as a generic Seyfert based on the line flux ratios in the spectrum.
To add further weight to the Seyfert classification, we can note that the X-ray luminosity of NED02 is between 10$^{42}$ and 10$^{43}$ erg s$^{-1}$.
This places the NED02 X-ray luminosity in the range of other local Seyferts (e.g. 
\cite{2017FrASS...4...34M}). The power law spectral indices and Nh values 
also are typical of AGNs as well (see e.g. Vasylenko et al. 2015).
From a qualitative point of view, the optical spectrum of NED02 is also very different from type I AGNs. It exhibits narrow emission lines contrary
to e.g. type I AGN detected in Abell 2163 
by \citeauthor{2008A&A...481..593M} (2008: see their Fig 1).

These optical spectroscopic data, despite covering a large wavelength domain and allowing detection of several spectral lines, are however limited by their low resolution: R$\sim$750. They are therefore not adapted to compute velocity dispersions. Moreover, they are not covering 100$\%$ of the galaxy area. In order to bypass these limitations, we got integral field spectroscopic data, on a very small spectral domain, strategically selected around the H$\alpha$/[NII] emission lines, that we describe in the next subsection.

\subsection{GHASP observations}

High-resolution Integral Field Unit (IFU) spectroscopy was
obtained with the Fabry-Perot instrument GHASP used on the OHP 193cm telescope on the 20/03/2023 with a $\sim$3arcsec seeing. GHASP has a field of view of 
5.8 × 5.8 arcmin$^2$, and is coupled with a 512 × 512 Image Photon 
Counting System (IPCS) with pixel size of 0.68 × 0.68 arcsec$^2$. It has 
a spectral resolution depending on the Fabry-Perot interferometer chosen, here R$\sim$11 000, see \cite{2002PASP..114.1043G}.
The full spectral range of 492 km/s of the Fabry-Perot at the redshifted
wavelength of the galaxy was scanned through 48
channels. The data reduction procedure adopted
to reduce the GHASP data have been extensively described in
 \cite{2008MNRAS.390..466E} and \cite{2019A&A...631A..71G}.

\begin{figure}[ht]
\includegraphics[width=9.cm,angle=0]{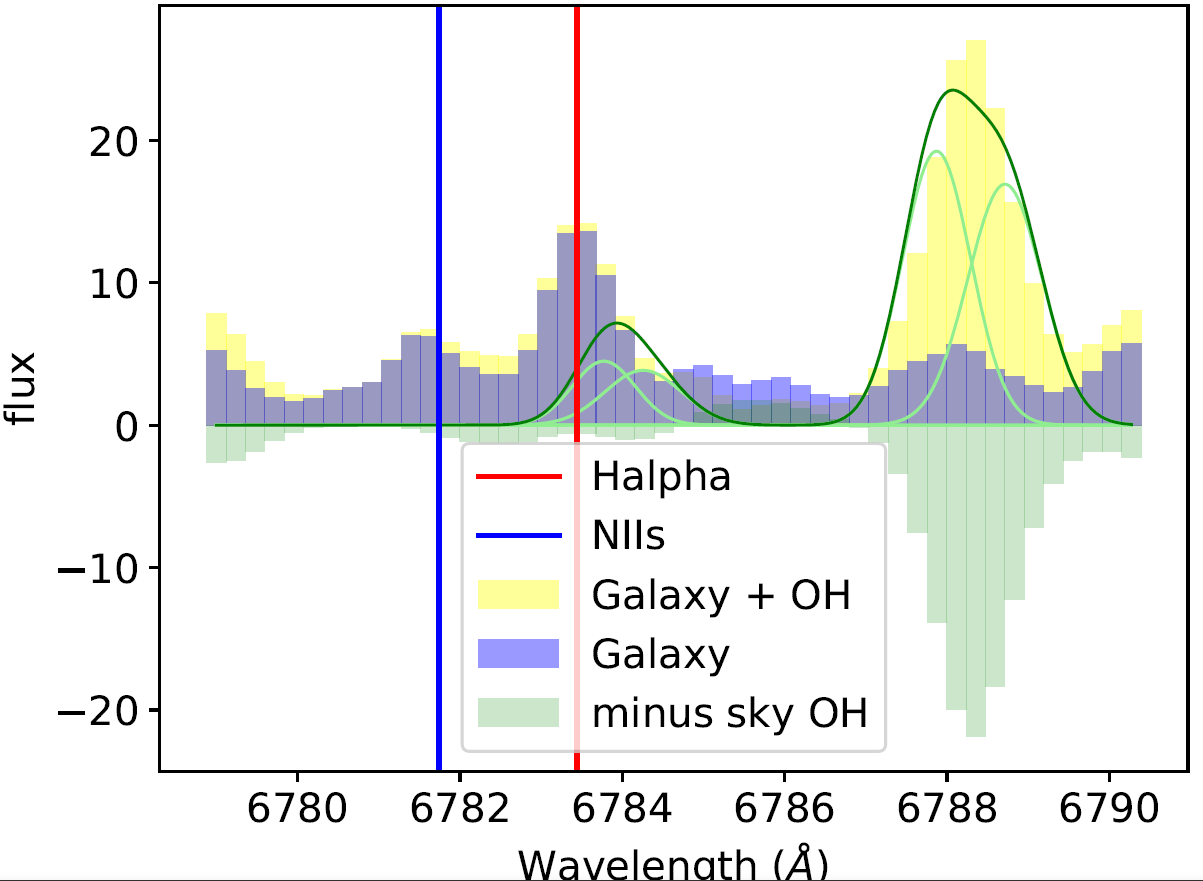}
\caption{\label{fig:PA1} 
 GHASP high resolution spectrum of the H$\alpha$/[NII] region measured on an area of  $\sim 4.6\,$arcsec$^2$ centered on the nucleus of the galaxy. Flux is in arbitrary units. The light green Gaussian curves are the expected individual OH night sky emission lines convolved at the Line Spread Function of the instrument ($\sim$ 10.7 km/s). The dark green curve is the sum of the light green lines. The mirrored light green histogram is the galaxy-free OH night sky lines measured on a region of $\sim 3.2\,$arcmin$^2$ around the galaxy. The net galaxy emission is given by the pink-blue histogram. Vertical blue and red lines are the expected redshifted location of the [NII]@6584 and H$\alpha$ respectively. Due to the cyclic nature of the Fabry-Perot interferometer, the redshifted [NII]@6584 is shifted by minus two free spectral ranges (one FSR $\sim$ 11.51\AA) to the blue and thus observed at $\sim$ -2 \AA\ from the H$\alpha$ line.}
\end{figure}

Gas central velocity dispersion of NED02 was estimated from GHASP data (Fig.~\ref{fig:PA1}) to be 11.2$\pm$1 km/s (tentatively corrected from instrument spectral resolution and based on the H$\alpha$ line), or 15.5$\pm$1 km/s not corrected.  The velocity dispersion has been measured considering that the H$\alpha$ emission can be modelled by a Gaussian function in dividing the measured FWHM by $2 \sqrt{2 \ln 2}$. The measure was made within an area of $\sim$4.6 $\arcsec$$^2$, equivalent to $\sim$2.2 kpc$^2$ at the redshift of the galaxy pair.

We used \cite{2022MNRAS.512.1765O} to 
statistically translate this value into a stellar velocity dispersion estimate of 34$\pm$1 km/s (corrected) and 41$\pm$1 km/s (not corrected). 
This can be directly related to the mass of the central black hole. \cite{2002ApJ...574..740T} would give $M_{\mathrm{BH}}=1.2\times10^{5}~\mathrm{M}_{\odot}$ (corrected), or $M_{\mathrm{BH}}=2.6\times10^{5}~\mathrm{M}_{\odot}$ (not corrected). Given the uncertainties in the relations 
of \cite{2022MNRAS.512.1765O} and \cite{2002ApJ...574..740T}, and our own uncertainty on the gas central velocity dispersion, the upper value of
the mass is: $M_{\mathrm{BH}}=5.3\times10^{5}~\mathrm{M}_{\odot}$. This value corresponds to the lower limit we estimated in \ref{MBH}, based on the optical variability of the core of NED02 (see below).

Finally, no significant H$\alpha$ emission was detected outside of the NED02 core. Some signal may be present to the northern outskirts of the galaxy, northern of MISTRAL extraction region \#5, but noise is too high to make this potential detection significant.

\section{Dynamical status of the CGCG 077-102 pair in A2063}
\label{sec:dynamical}
\begin{figure}[ht]
\includegraphics[width=9.cm,angle=0]{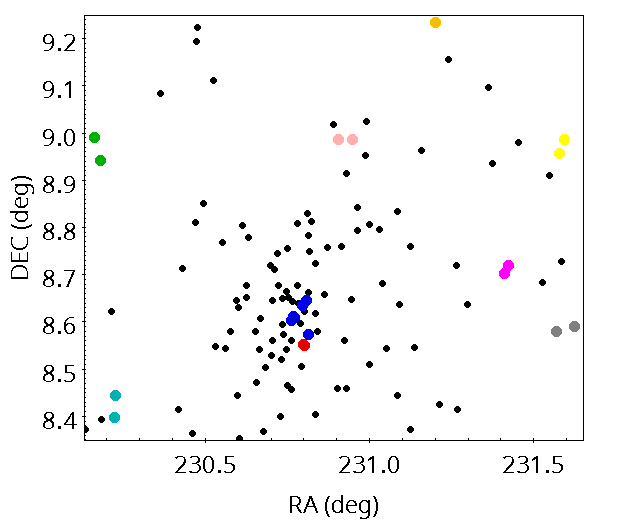}
\includegraphics[width=9.cm,angle=0]{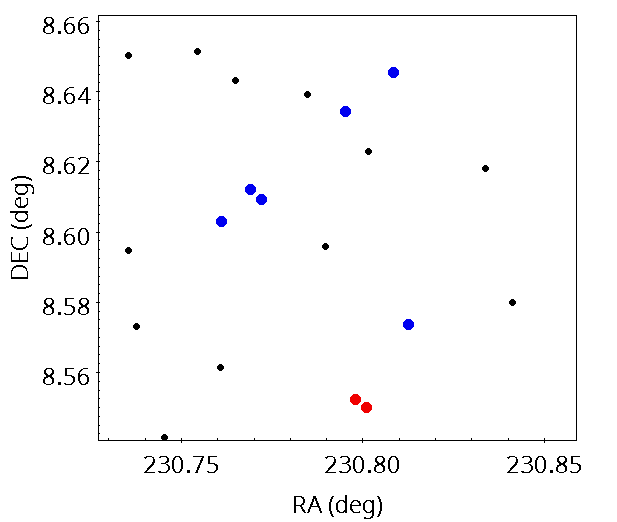}
\caption{\label{fig:n1} Sky distribution (within the A2063 field of view) of the galaxies with a measured spectroscopic
redshift (from NED) between 0.0285 and 0.03963 (black dots). The cluster bulk (as determined by the SG method) is shown as
blue filled circles. The CGCG 077-102 pair is the red filled circles. Other SG-detected pairs are other filled circles with
different colors. Top: full field of view. Bottom: zoom on the cluster center.}
\end{figure}

We now investigate the dynamical status of the CGCG 077-102 pair in A2063 in order to detect possible large scale-induced
phenomenon in the NED02 AGN behaviour. 
The spatial distribution of galaxies within the A2063 field of view (with available spectroscopic redshifts in NED) is shown in Fig.~\ref{fig:n1}. This catalogue includes 177 galaxies with spectroscopic redshifts. 

It was selected to be included in a 1deg radius from the cluster center. This represents a physical 
$\sim$2.5 Mpc radius. A2063 has a measured X-ray temperature of $\sim$4.1 keV \cite[]{1998MNRAS.301..881E}. This translates into a total mass of 4 10$^{-14}$ E(z) M$_{500}$ h$_{70}^{-1}$ \cite[]{2016A&A...592A...4L}, equivalent to a radius of $\sim$1.5 Mpc. Our catalog is therefore covering all the densest part of the cluster.

In order to also select the densest cluster parts along the redshift axes, we chose to select the redshift range where the velocity gap between two consecutive galaxies along the line of sight was larger than 200 km s$^{-1}$ \cite[for a similar technique]{1998A&A...331..439A}. This leads to cut the redshift range between \z =0.0285 and \z =0.03963. This corresponds approximately to $\pm$2.6 cluster velocity dispersions (estimated to be 639 km s$^{-1}$, see below).

We searched for substructures by appling the Serna-Gerbal method
\cite[SG hereafter]{1996A&A...309...65S}
to the velocity catalog. This hierarchical method allows cluster substructures to be
extracted from
a catalog containing positions, magnitudes, and velocities, based on the calculation of their relative negative binding
energies. The method produces a list of galaxies which belong to the selected substructures, as well as information on their
binding energy. We used a M/L ratio in the V-band of 100. We checked that using M/L ratios between 100 and 400 does not induce
the detection (or the absence of detection) of additional cluster substructures. 

\begin{figure}[ht]
\includegraphics[width=9.cm,angle=0]{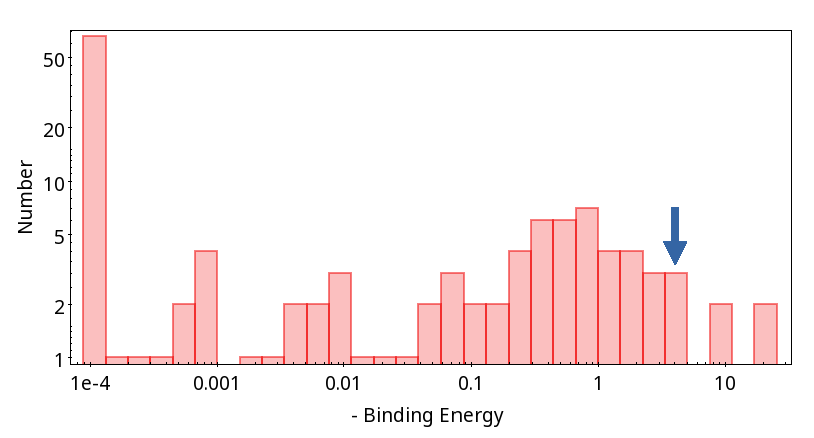}
\caption{\label{fig:n2} 
Binding energy histogram obtained by applying the SG technique for the 177 galaxies shown in Fig.~\ref{fig:n1}. The x-axis is the relative binding energy (in arbitrary units). The y-axis shows the number of galaxies for a given binding energy. The CGCG 077-102
pair is shown as the vertical arrow.
}
\end{figure}

The SG method reveals the existence of a
main bulk (the most tightly bound galaxies inside the cluster potential well, assumed to constitute the cluster seed). The galaxies in this bulk have a typical extension on the sky of $\sim$50 kpc (see Fig.~\ref{fig:n1}) and their velocity dispersion is 317 km/s. When considering all the galaxies in the cluster, the velocity dispersion is
639 km.s$^{-1}$. 
In addition, eight galaxy dynamical-pairs are also detected. Their positions are displayed in Fig.~\ref{fig:n1}. The most cluster-potential
bound pair is constituted by the presently studied pair (CGCG 077-102, NED01 and NED02). Less than 5$\%$ of the known cluster
galaxy members are more linked to the cluster potential (see Fig.~\ref{fig:n2}).
Moreover, Fig.~\ref{fig:n1} shows that the CGCG 077-102 pair is very centrally located inside the cluster, contrary to all other pairs
which are detected inside the cluster outskirts.




This clearly means that the CGCG 077-102 pair is not a recent member of the A2063 and has had enough time to fall deep
inside the cluster potential. The pair underwent the cluster influence for a long time.

Our MISTRAL optical follow-up was however not regular enough to allow a study of the flux variability of the NED02 core. We therefore used the Zwicky Transient Facility (ZTF) for this purpose in the next section.

\section{Optical light curve analysis of CGCG 077-102 NED02}
\label{sec:optical}
In many works, the flux or magnitude variability is often used to complement the systematic identification of AGNs in samples with a large amount of data, such as multi-epoch surveys \citep[e.g.,][]{2008Trevese,2012Young,2015DeCicco,2020Poulain,2022Burke}. The variability is an important characteristic observed in all wavelength ranges of the electromagnetic emission of AGNs as it can be related to physical processes that occur in their accretion disk and jets \citep[see][]{2017Padovani}. However, the mechanisms behind it still need to be better understood. In this section, we describe the datasets and the time-series optical analysis of the galaxy NED02. We aim to quantify the optical variability associated with the AGN activity as well as to search for a periodic (or quasi-periodic) behavior in the light curves to understand the mechanisms behind the interactions of the galaxy pair. 

\subsection{Zwicky Transient Facility DR15}

We used public data available on the 15th data release of the Zwicky Transient Facility (ZTF), a time-domain survey that uses the 1.22~m Samuel Oschin Schmidt telescope at the Palomar Observatory and a 47 deg² field-of-view camera to cover the northern sky in $r$, $g$, and $i$ optical filters \citep{Bellm2019,Graham2019}, in order to obtain four years of observations from NED02 and its companion NED01. All details about the data-processing pipeline (ZTF Science Data System, ZSDS) are described in \cite{Masci2019}.



We selected the nearest sources inside 5\arcsec ~from the optical center of the galaxies on ZTF, which correspond to light curves in $r$, $i$, and $g$ filters. As the data of each filter and CCD-quadrant are processed separately by the ZTF's pipeline, we found multiple light curves in the same filter for a single object. However, the combination of these multiple light curves can introduce spurious variability due to the calibration differences \citep{vanRoestel_2021}. For this reason, we chose the light curves with the highest number of observations to study. Also, due to the small number of observations, we excluded the $i$ band light curve from our analysis (see Table \ref{tab:ZTFsources} for details).

\begin{table*}[ht]
\centering
\caption[]{The four nearest sources from the optical center of NED02 and NED01.}
\label{tab:ZTFsources}
\begin{tabular}{ccccc}
\hline
Galaxy  & oid$^{(a)}$ & Filter   & Nobs$^{(b)}$  & Selected \\ \hline
\multirow{4}{*}{\makecell{CGCG 077-102 \\ NED02}} 
& 531302400031665  & $i$ & 91   & No       \\
& 531202400030069  & $r$ & 422  & Yes      \\
& 1526110300008580 & $g$ & 11   & No       \\
& 531102400013043  & $g$ & 300  & Yes      \\
\hline
\multirow{4}{*}{\makecell{CGCG 077-102 \\ NED01}} 
& 531102400008167  & $g$       & 300  & Yes      \\
& 531202400023741  &  $r$      & 422  & Yes      \\
& 1526110300018677 & $g$    & 11   & No       \\
& 531302400029994  & $i$    & 91   & No       \\ \hline
\multicolumn{5}{l}{\footnotesize$^{(a)}$ The object identifier of ZTF.}\\
\multicolumn{5}{l}{\footnotesize$^{(b)}$ Number of public observations with high image quality flag.}\\
\end{tabular}
\end{table*}

In addition, we obtained all light curves through the ZTF application program interface\footnote{\url{https://irsa.ipac.caltech.edu/docs/program_interface/ztf_lightcurve_api.html}}, where we applied a filter in a search parameter (\texttt{BAD\_CATFLAGS\_MASK=32768}) to select only observation epochs with high image quality flags (i.e., \texttt{catflags=0}). This criterium excludes observation epochs affected by clouds and/or the moon, as described in ZTF documentation\footnote{\url{https://irsa.ipac.caltech.edu/data/ZTF/docs/releases/ztf_release_notes_latest}}. 

\subsection{Noise removing}   
We noticed some similarities between the light curves of NED02 and other galaxies in the neighborhood of the A2063 cluster by visual inspection, suggesting the existence of an unknown source of noise affecting extended objects at this observed field on the same epochs.

\begin{figure}[ht]
    \centering
    \includegraphics[width=9.12cm]{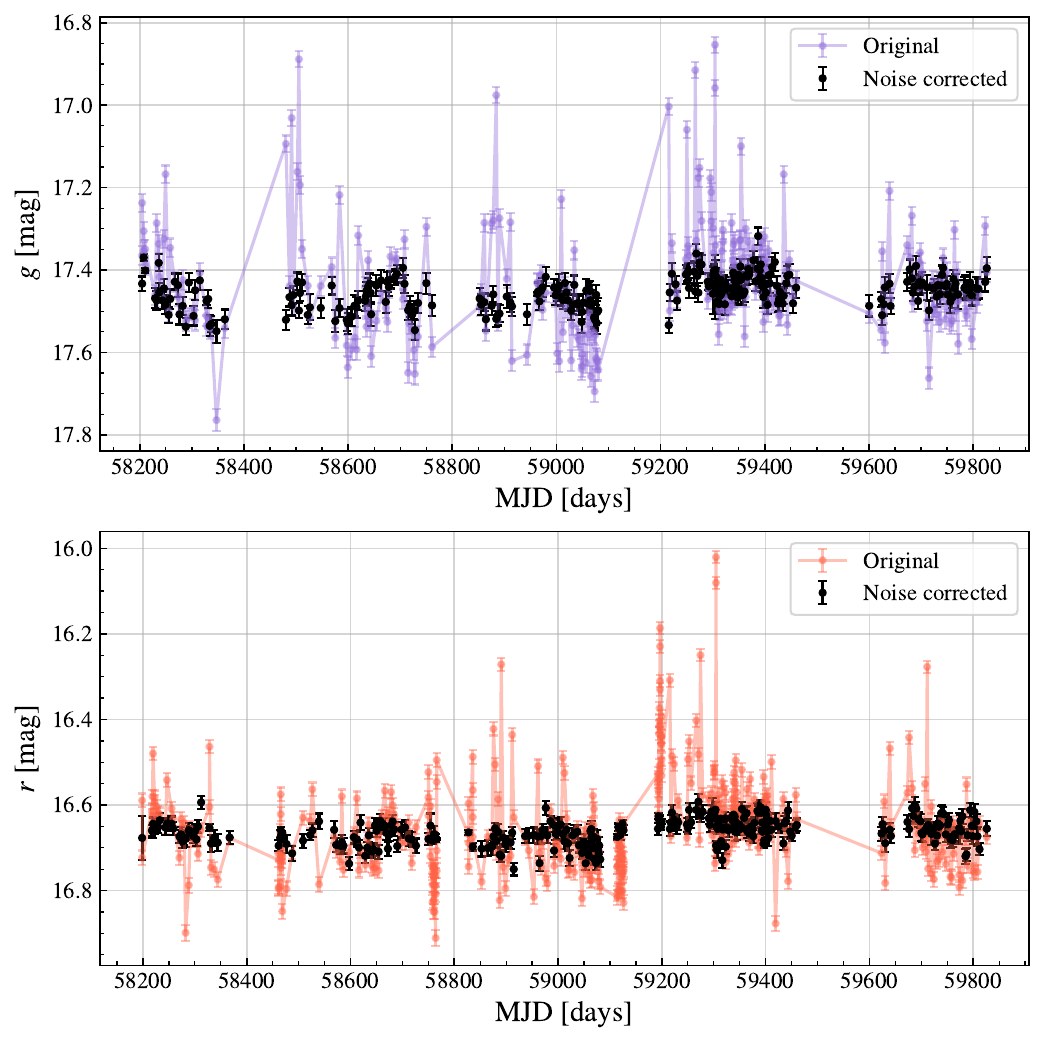}
    \caption[]{Comparison between the light curves in $g$ (top panel) and $r$ (bottom panel) bands of NED02 before and after the noise correction with PDT and calculation of the intranight observations mean. The purple and orange connected dots represent the original light curves in their respective filters ($g$ and $r$), and the black dots with error bars represent the noise-corrected light curve.}
    \label{fig:NED02_detrended}
\end{figure}


The presence of systematic and random trends in light curves may be associated with atmospheric events (such as cloud passages and air masses), instrumental effects (CCD noise and telescope vibration), and imperfections in the data reduction (e.g., flat fielding). This can introduce false positives, leading us to misinterpret the results (e.g., false detection of periodic or quasi-periodic signals in AGNs). Several works have also reported similar problems in the light curves of their targets \citep[e.g.,][]{2006Pont,2010Kim,2010Mislis,2013Gibson}.

Several methods and tools to improve the signal-to-noise of the light curves are available in the literature, such as de-trending algorithms \citep[e.g.,][]{2005Kovacs,2005Tamuz,2009Kim}, smoother filters 
\citep[e.g.,][]{1964Savitzky}, and Bayesian methods \citep[e.g.,][]{2020Taaki}. In this work, we used the Photometric DeTrending Algorithm Using Machine Learning\footnote{\url{https://github.com/dwkim78/pdtrend}} \citep[PDT,][]{2009Kim}, which applies the balanced iterative reducing and clustering using hierarchies algorithm (BIRCH, \citealt{Zhang1996}) in its latest version to identify common trends in a set of light curves used as templates. After that, the algorithm constructs master trends that will be used to remove trends from individual light curves (see \citealt{2009Kim} for more details).

We must consider some aspects before selecting a set of light curves used as templates by PDT.
\begin{itemize}
    \item The light curves must be synchronized in time.

    \item We have to select a template set of bright objects that are not saturated.

    \item It is recommended to choose non-variable objects (or, at least, a high number of them) in the same field as your target since the algorithm is sensitive to the choice of light curves \citep{2009Kim}.
    
    \item The PDT's manual recommends having a template set of at least 50 light curves to run the code.
    
\end{itemize}

However, the sample selection process becomes challenging since there are not enough objects that follow these requirements in the same field as NED02 and that have data available on ZTF in all filters used. So, instead, we selected the galaxy companion NED01, which is a passive galaxy, as described in Section \ref{sec:spec}. By definition, the galaxy pair was observed in the same epochs, in the same weather conditions, and was susceptible to the same instrumental errors. Also, due to the passive state of NED01, we do not expect variability in its light curve related to physical processes on the SMBH. Therefore, we assumed it could be used as a light curve reference to subtract the noise in the galaxy NED02 since the PDT detects similarities between light curves.

We ran the code using a template set composed of NED01 and NED02 light curves to find master trends. After that, we detrended the NED02 light curve. The error propagation was computed using a Monte Carlo method 
\citep[see][]{jcgm2008}\footnote{\url{https://www.bipm.org/en/committees/jc/jcgm/publications}}, as described below: 
\begin{enumerate}
    \item We added noise to the original data to simulate a sample of 50 new light curves.
    \begin{equation}
        m = m_{0} + \sigma_{0} \cdot \sigma_{noise},
    \end{equation}
    where $m$ is the detrended magnitude, $m_{0}$ is the magnitude of the non-detrended light curve, $\sigma_{0}$ is the uncertainty of $m_{0}$, and $\sigma_{noise}$ is the random noise generated from a normal distribution of mean 0 and variance 1. 

    \item After detrending each simulated light curve with the same master trends computed before by the PDT, we calculated the standard deviation of each observation epoch to get the new uncertainty.

\item Due to a significant spread in the magnitudes measured in the same night, the final step in our time series analysis was calculating the mean of intranight observations. The light curves before and after the noise correction are shown in Figure \ref{fig:NED02_detrended}. 
\end{enumerate}
Finally, we repeated the last three steps (detrending, error propagation, and mean of the intranight observations) for the NED01 light curves only for comparison. The results are shown in Appendix \ref{app:ned01_lc}.

\subsection{Optical Variability}
\subsubsection{Excess Variance}

\begin{table*}[ht]
\centering
\caption[]{Statistical values of the NED02 light curves before and after noise correction in both filters ($g$ and $r$).}
\begin{tabular}{ccccc}
\hline
         & \multicolumn{2}{c}{ZTF $g$-band} & \multicolumn{2}{c}{ZTF $r$-band} \\ 
\hline
Parameter       & \begin{tabular}[c]{@{}c@{}}Original\\ light curve\end{tabular} & \begin{tabular}[c]{@{}c@{}}Noise\\ corrected\end{tabular} & \begin{tabular}[c]{@{}c@{}}Original\\ light curve\end{tabular} & \begin{tabular}[c]{@{}c@{}}Noise\\ corrected\end{tabular}  \\ \hline
$m_{mean}$$^{(a)}$ & 17.43 $\pm$ 0.01 & 17.454 $\pm$ 0.002 & 16.65 $\pm$ 0.01 & 16.662  $\pm$ 0.002 \\
$\sigma^{2}_{\mathrm{XS}}$ & 0.017 $\pm$ 0.001  & 0.0009 $\pm$ 0.0001 & 0.013 $\pm$ 0.001 & 0.0006$\pm$ 0.0001 \\
$m_{max}$$^{(b)}$         & 17.76  & 17.55 & 16.91  & 16.75        \\
$m_{min}$$^{(c)}$          & 16.85  & 17.32  & 16.02   & 16.59  \\
$m_{(1-3)}$$^{(d)}$&17.46 $\pm$ 0.01&17.473 $\pm$ 0.003&16.69 $\pm$ 0.01&16.677 $\pm$ 0.002\\
$m_{(4-5)}$$^{(d)}$&17.41 $\pm$ 0.01&17.437 $\pm$ 0.003&16.60 $\pm$ 0.01&16.647 $\pm$ 0.002 \\
Skewness        & -1.3 & 0.04   & -1.4  & 0.25 \\
 \hline
\multicolumn{5}{l}{\footnotesize$^{(a)}$ Mean value of the light curve.}\\
\multicolumn{5}{l}{\footnotesize$^{(b)}$ Maximum magnitude.}\\
\multicolumn{5}{l}{\footnotesize$^{(c)}$ Minimum magnitude.}\\
\multicolumn{5}{l}{\footnotesize$^{(d)}$ Mean level of the observed cycles interval of the light curve in units of magnitude.}\\
\multicolumn{5}{l}{\footnotesize Note -- the parameters ($m_{mean}$, $\sigma^{2}_{\mathrm{XS}}$, $m_{max}$, and $m_{min}$) are in units of magnitude.}
\end{tabular}
\label{tab:stat_ned02}
\end{table*}


The normalized excess variance, $\sigma_{\mathrm{NXS}}^{2}$, introduced by \cite{1997Nandra}, measures the intrinsic variability of an object due to the existence of measurement errors that will contribute to an "extra variance" in the data \citep{2003Vaughan}. In this work, we used the unnormalized version of the excess variance in magnitude units to quantify the variability of the light curves, which is defined as
\begin{equation}
    \sigma^{2}_{\mathrm{XS}} =  S^{2} - \left \langle \sigma^{2}_{\mathrm{err}} \right \rangle,
\end{equation}
where $S^{2}$ is the variance of the light curve and $\left \langle \sigma^{2}_{\mathrm{err}} \right \rangle$ is the mean square error:
\begin{equation}
    \left \langle \sigma^{2}_{\mathrm{err}} \right \rangle = \frac{1}{N_{\mathrm{obs}}} \sum^{N_{\mathrm{obs}}}_{i=1} \sigma_{\mathrm{err},i}^{2},
\end{equation}
where $N_{\mathrm{obs}}$ is the number of observations and $\sigma_{\mathrm{err},i}$ is the uncertainty of the individual magnitude $i$. Following \cite{2022Yuk}, the error in the excess variance is
\begin{equation}
    \Delta (\sigma^{2}_{\mathrm{XS}}) = \sqrt{\frac{2}{N_{\mathrm{obs}}-1}}S^{2}.
\end{equation}


The excess variance is often used as a selection parameter to determine if the object has a significant variability to be considered as an AGN of a certain class \citep[e.g.,][]{2022Yuk,2023LopezNavas}. In addition, several authors have studied its correlation with physical properties, such as the $\sigma_{\mathrm{NXS}}^{2}-M_{\mathrm{BH}}$ correlation in X-ray data \citep[e.g.,][]{2012Ponti,2010Zhou}, which makes this parameter important to be estimated \citep[see][]{2017Padovani}.

\subsubsection{Variability Results}
\label{subsub:var}

\begin{figure}[ht]
    \centering
    \includegraphics[width=9.3cm]{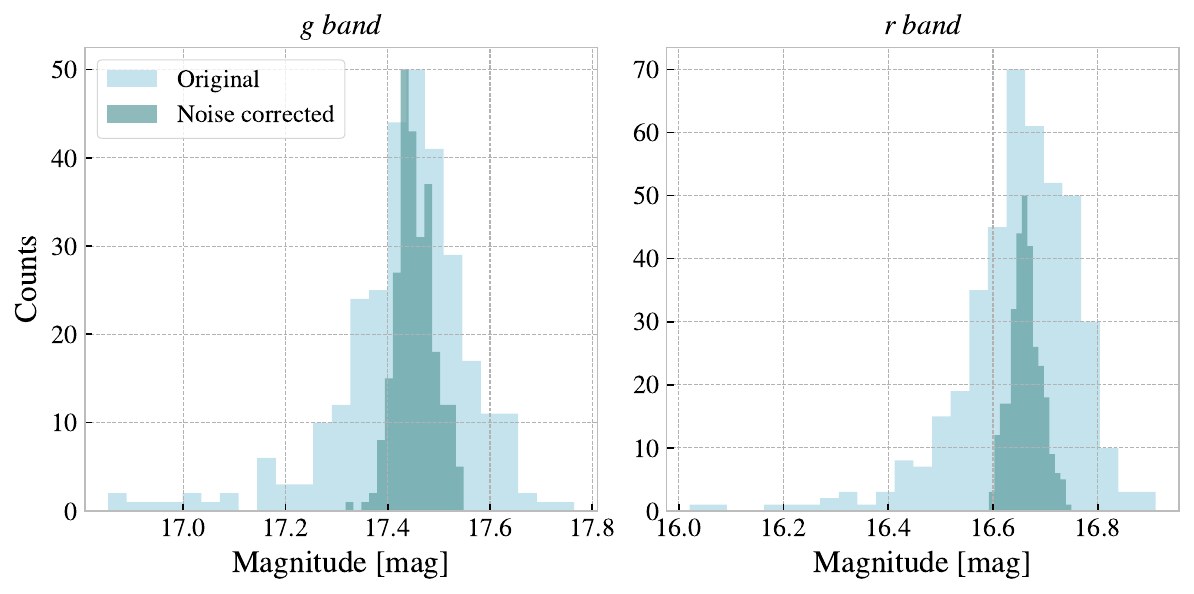}
    \caption[]{Histograms of magnitude distribution before (light blue) and after (dark turquoise) the noise correction of NED02 light curves in $g$ (left) and $r$ (right) bands.}
    \label{fig:NED02_hist}
\end{figure}

The histograms of magnitude distribution before and after noise correction are shown in Figure \ref{fig:NED02_hist}. First, we observed that the light curves magnitude distribution became more compact around the mean value after the noise correction, i.e., the observed variability decreased. On the other hand, the magnitudes are widely spread before the correction, and its distribution was asymmetric, showing a tail on the left side in the direction of low magnitudes (negative value for skewness, see Table \ref{tab:stat_ned02}). However, it is likely a consequence of the presence of outliers, and we can confirm this with more observations to check if the pattern will repeat itself. The statistical results of the NED02 light curve are given in Table \ref{tab:stat_ned02}.

We noticed a difference in the mean level value of the first three and last two cycles\footnote{We define a \textit{cycle of observations} as an interval of quasi-sequential data surrounded by a long period without observations. For example, the light curves used have five cycles of observations.} of the light curves observations in both filters (see Table \ref{tab:stat_ned02}). This behavior is present even before the detrending process of the light curves and seems to happen in both filters quasi-simultaneously (Figure \ref{fig:NED02_detrended}), indicating it could not be an artifact introduced by the PDT code. Thus, new observation cycles will help us understand the nature of this very long-timescale variability pattern. If the very long-term variability (from years to decades) is true, then a small excess variance (or normalized excess variance) is expected because it is more sensitive to short-term variability, and most of the observing magnitudes are around the light curve mean until now. 

We did not detect a short-term variability on time scales from weeks to months in any filter due to the small excess variance (Table \ref{tab:stat_ned02}). We cannot exclude the possibility of signal amplitude over-subtraction during the detrending process because we expected a nearly constant profile for the NED01 light curve, but it seems to have some artifacts (see Appendix \ref{app:ned01_lc}). Despite the small excess variance (before and after the noise correction), it is still more significant in the $g$-band than in the $r$-band. The variability on timescales from months to years in the detrended light curves is also more significant in the $g$-band (see Figure \ref{fig:NED02_detrended}). Regarding the long-term variability, a detailed discussion is described in Subsection \ref{subsec:periodicity}. Furthermore, several studies have detected larger variability amplitudes on time scales from months to years in optical bands than in the X-rays, while on time scales from hours to days, the variability observed in the X-rays is larger than in the optical \citep[][and references therein]{2008Arevalo,2009Arevalo,2003Uttley}. In general, the challenging task of distinguishing the lower luminosity AGNs from the normal galaxies in the optical wavelengths is evident due to the observed low optical variability \citep[see also][]{2017Padovani}. Therefore, the optical variability as a unique criterium for AGN selection might be insufficient.

In the future, we will improve the techniques for noise removal to avoid biases and misinterpretation of the results through our requested optical telescope time to perform intra-night observations, which will help us identify outliers, to better estimate the noise, and confirm whether NED02 does not have a very short variability (on timescales from hours to a few days).

\subsection{Periodicity: Damped Random Walk}
\label{MBH}

In this section, we describe one of two methods used to model and characterize the long-term oscillatory behavior observed in the g-band light curve of NED02. We did not analyze the $r$-band to avoid the wrong estimate of constraints due to its low variability (Subsection \ref{subsub:var}) and the non-existence of possible periodic trends that can be distinguished from the likely unknown artifacts introduced during the detrending process without any bias, as shown in Figures \ref{fig:NED02_detrended} and \ref{fig:NED02_r_detrended}. A detailed description of the second method, the Lomb-Scargle periodogram analysis, can be found in Appendix \ref{ap:LSP}.

\begin{figure*}[ht]
    \centering
    \includegraphics[scale=0.8]{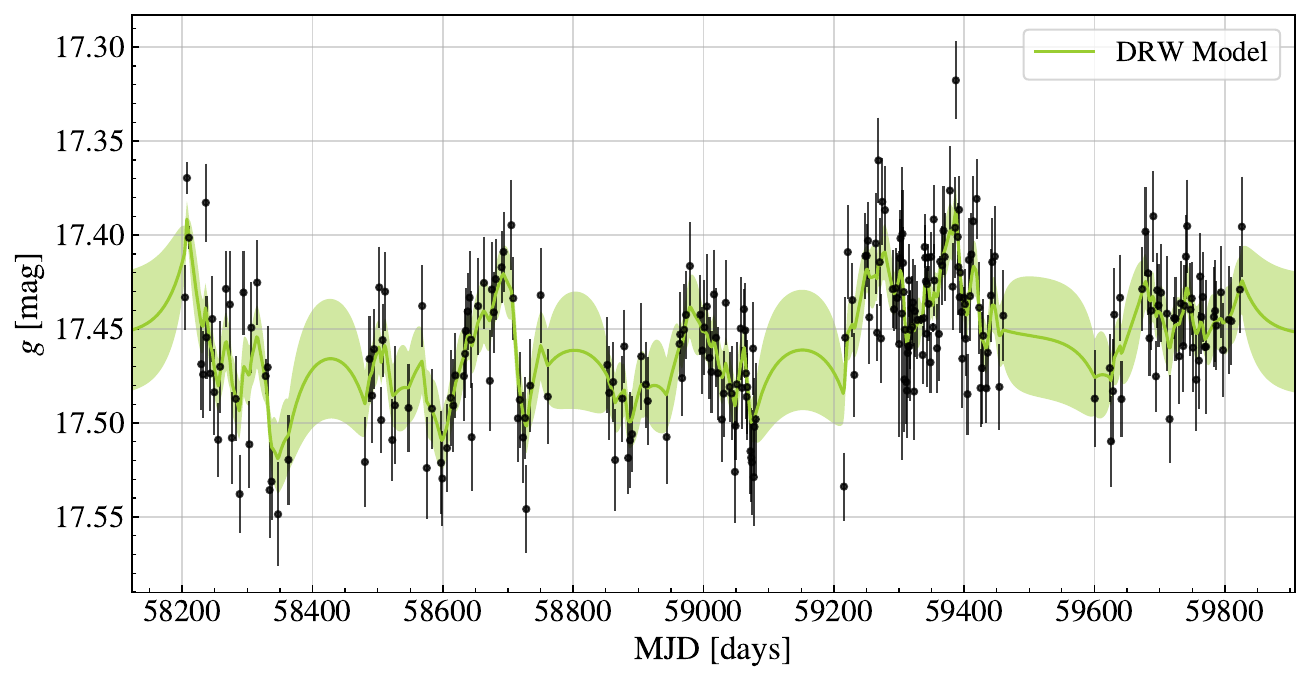}
    \caption[]{The detrended light curve of NED02 in $g$-band (black dots). The best-fitted DRW model with a $1\sigma$ of uncertainty (shaded area) is shown in green. The $1\sigma$ of uncertainty is the 16th and 84th percentiles of the MCMC sample.}
    \label{fig:NED02_DRW}
\end{figure*}  

The Damped Random Walk (DRW) is a stochastic model often used to describe the accretion disk variability of AGNs \citep{2009Kelly}. It is the lowest-order model of the continuous-time autoregressive moving average (CARMA) models introduced by \cite{2014Kelly} \citep[see][for a review]{2019Moreno}. The DRW model gives a PSD characterized by a white noise slope ($\propto f^{0}$) at low frequencies and a red noise slope ($\propto f^{-2}$) at high frequencies, where the transition between these regions indicates a frequency related to a characteristic damping timescale, $\tau_{\mathrm{DRW}}=(2\pi f_{\mathrm{DRW}})^{-1}$, that is correlated with black hole mass \citep{2009Kelly,2021Burke}.

In this work, we use the \texttt{Celerite} \citep{2017celerite} Python package to implement the Gaussian Process (GP) regression and to build the DRW kernel function with an additional white noise term (also called jitter term) to model the NED02 light curve. Then, we use the \texttt{emcee} \citep{2013ForemanMackey} package to implement the Markov Chain Monte Carlo (MCMC) sampler algorithm to analyze the posterior probability distribution, assuming uniform priors for all free parameters. For all parameters, we computed the median of the MCMC samples and their 16th and 84th percentiles for the uncertainties. The covariance function (kernel) of the DRW with an additional jitter term is given by
\begin{equation}
    k(t_{nm})=2\sigma^{2}_{\mathrm{DRW}}\exp(-t_{nm}/\tau_{\mathrm{DRW}}) + \sigma_{\mathrm{WN}}^{2} \delta_{nm},
\end{equation}
where $\sigma_{\mathrm{DRW}}$ is the amplitude term, $t_{nm}=\left |t_{n} -t_{m}  \right |$ is the time lag between the measurements $t_{m}$ and $t_{n}$, $\sigma_{\mathrm{WN}}$ is the amplitude of the white noise signal, and $\delta_{nm}$ is the Kronecker delta function.

\begin{figure}[ht]
    \centering
    \includegraphics[scale=0.49]{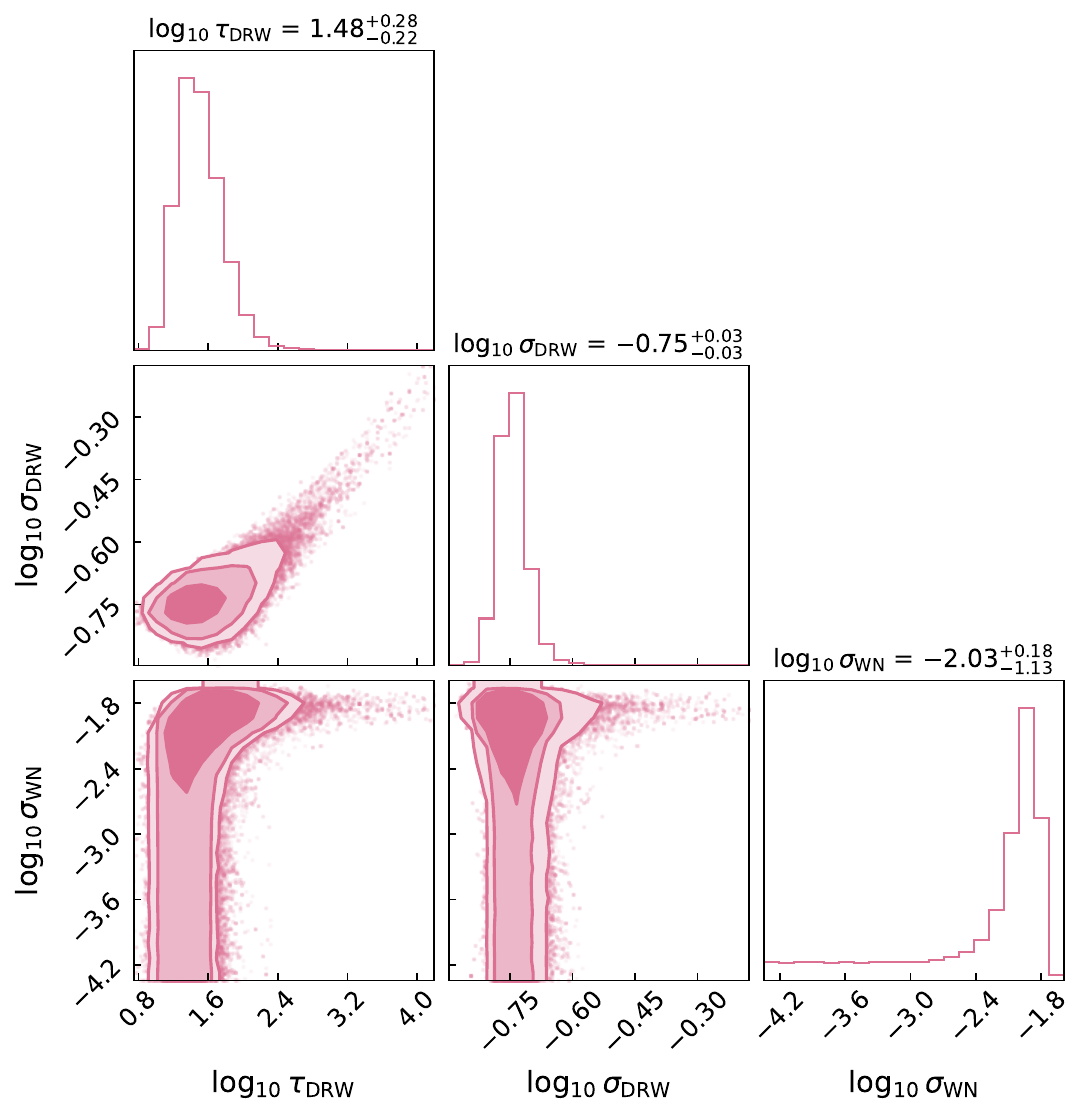}
    \caption[]{Posterior probability distributions of the parameters from the DRW and jitter kernels. The contours show 68\%, 95\%, and 99\% confidence intervals, and the pink dots are the MCMC samples. The parameter $\tau_{\mathrm{DRW}}$ has units of days, and the amplitudes ($\sigma_{\mathrm{DRW}}$ and $\sigma_{\mathrm{WN}}$) have units of magnitude. The parameter uncertainties were computed using the 16th and 84th percentiles of the MCMC samples.}
    \label{fig:NED02_DRW_corner}
\end{figure}

\begin{figure}[ht]
    \centering
    \includegraphics[scale=0.63]{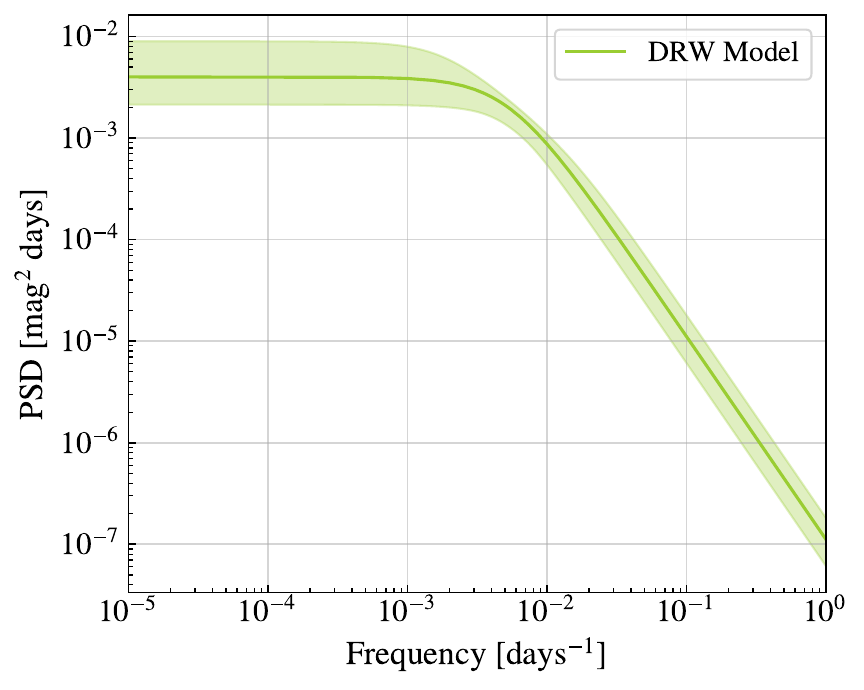}
    \caption[]{The power spectrum density of the best-fitted DRW model obtained with \texttt{Celerite} (solid green line). The green shaded area shows the region between the 16th and 84th percentiles. The characteristic frequency that defines the transition between the white and red noise in the PSD is $f_{\mathrm{DRW}}= 0.0053_{-0.0025}^{+0.0034}$ days$^{-1}$.}
    \label{fig:NED02_DRW_PSD}
\end{figure}

The best-fitted DRW model that describes the entire detrended light curve of NED02 (Figure \ref{fig:NED02_DRW}) gives the parameters $\tau_{\mathrm{DRW}}=30_{-12}^{+28}$ days, $\sigma_{\mathrm{DRW}}=0.18^{+0.01}_{-0.01}$ mag, and $\sigma_{\mathrm{WN}}=0.009_{-0.009}^{+0.005}$ mag. The posterior probability distributions of the fitted parameters are shown in Figure \ref{fig:NED02_DRW_corner}, and the power spectrum density of the DRW model is shown in Figure \ref{fig:NED02_DRW_PSD}.

We computed the supermassive black hole mass constraints using the $\tau_{\mathrm{DRW}}-M_{\mathrm{BH}}$ scale relations available in the literature in order to compare it with the one we obtained from the velocity dispersion measures of the optical spectra. Following \cite{2009Kelly}, we found $M_{\mathrm{BH}}=10^{6.73_{-0.39}^{+0.51}}~\mathrm{M}_{\odot}$. From \cite{2021Burke}, we have $M_{\mathrm{BH}}=10^{6.65_{-0.55}^{+0.72}}~\mathrm{M}_{\odot}$, and $M_{\mathrm{BH}}=10^{6.43_{-0.70}^{+0.91}}~\mathrm{M}_{\odot}$ for the \cite{2023Wang} relation. The transitional frequency, $f_{\mathrm{DRW}}= 0.0053_{-0.0025}^{+0.0034}$ days$^{-1}$, is not compatible with the $f_{break}$ computed in the Lomb-Scargle analysis (Appendix \ref{ap:LSP}). It is an expected result because we analyzed only a small region of the light curve in the Lomb-Scargle periodogram analysis. Also, the Lomb-Scargle method expects to model periodic signals, which is probably not the NED02 case.

Stochastic models usually give a good description of the well-sampled continuous AGN light curves. However, the DRW model can deviate from the real light curves in the presence of gappy data or limited length, introducing bias on the predicted parameters \citep{2017Kozlowski}. Furthermore, recent studies have shown that some AGN light curves require more flexible models \citep[e.g.,][]{2011Mushotzky,2013Zu,2015Kasliwal,2017Guo}, such as high-order CARMA models \citep{2014Kelly} that do not have a fixed PSD like the DRW \citep{2019Moreno}. For example, the variability driven by multiple processes in the accretion disk or other unknown perturbations can be responsible for deviations from the DRW model \citep{2013Zu,2017Dobrotka,2017Guo,2019Moreno}.

Here, we show that NED02 has an optical light curve better described by the DRW model than the classical Our mass method, as the other AGNs found in the literature. That means NED02 has aperiodic stochastic variability in long timescales. However, we must be careful with the estimated parameters due to all the potential issues related to the detrending process already mentioned before, degeneracy in DRW modeling \citep{2016Kozlowski}, and the possible bias introduced in the $\tau_{\mathrm{DRW}}-M_{\mathrm{BH}}$ relations found in the literature due to the restricted samples of these studies \citep{2021Burke}.

In future work, we will take advantage of our submitted proposals to get nearly synchronized optical and X-rays observations with MISTRAL and Chandra telescope. This will allow us to study the existence of time delays in the optical emission regarding the X-rays, as well as to compare their variability amplitudes in different timescales, which can shed some light on the scenarios that try to explain the physical processes responsible for the optical emission of the accretion disk. Moreover, more observation cycles will be available on the next ZTF data releases, allowing the increase of the NED02 light curve length. Consequently, we will be able to improve the accuracy of the parameter fitting with the DRW model. Also, we will verify whether more flexible high-order CARMA models are necessary to model the light curve due to the possible more complex processes that are only evident in very long timescales.

\section{Summary and Conclusions}

In the framework of investigating the link between central super massive black holes
in the core of galaxies and the galaxies themselves, we detected a variable X-ray source
in the CGCG 077-102 galaxy pair. In Tab. \ref{tab:IR} we display typical magnitudes of NED01 and NED02 drawn from the WISE catalog\footnote{This publication makes use of data products from the Wide-field Infrared Survey Explorer, which is a joint project of the University of California, Los Angeles, and the Jet Propulsion Laboratory/California Institute of Technology, funded by the National Aeronautics and Space Administration. }. Only one of the pair-members, NED02, harbors an AGN.
NED02 stands out also as being the only one present in the list in longer wavelengths in the Sptizer source catalog, e.g. the MIP 70$\mu$m band and in the X-rays (see Tab. \ref{tab:fitXspec}). Hence, literature searches for galaxy pairs with at least one AGN did not find the NED01-NED02 pair.

\begin{table*}[ht]
\caption[]{Summary of IR fluxes for NED02 and NED01 \label{tab:IR}}
\centering
\begin{tabular}{lccccc}
\hline
\hline
\hline
ID & WISE & J  &  H & K$_{\rm{s}}$ & WISE4 \\
\hline
NED01& J152311.51+083308.7& 14.832&14.216&13.944 & 9.087\\
\hline
NED02 & J152312.25+083259.1 &14.695  & 13.980& 13.530 & 6.534 \\
\hline
\hline
\end{tabular}
\end{table*}

Using \textit{Chandra} data and XMM-\textit{Newton} catalogue, we showed that there is no evidence of 
multiple components or extended component within the X-ray source. Only new on-axis \textit{Chandra} data 
could definitively confirm if there is an X-ray single source in the core of NED02. However, single X-ray spectral
models do not produce worse fittings than two-component models, and the X-ray source is within 0.2\arcsec\ of the core. Thus the X-ray emission is most likely coming from a single X-ray source, associated with a SMBH. 
We therefore considered but rejected a two X-ray source model.  First, because a two-component model was not needed to fit the X-ray spectrum and second because the position of the X-ray source is within error of the optical nucleus the galaxy.

We found that the 0.5--10\,keV flux varied between observations by factors up to $\sim$2 and this variation is likely non-random. We show a short term X-ray variability of $\sim$4~days and a long term variability lower than $\sim$700~days.
However we reject the possibility that we detected a TDE because no typical decay 
$t^{-5/3}$ was observed nor the typical black body spectrum with $kT$ near 0.1\,keV \cite[e.g.][]{2021ARA&A..59...21G}.

To characterize this source in optical, we also obtained low resolution optical spectroscopy in order to detect characteristic spectral lines over 
extended areas of the galaxy pair. NED01 only shows absorption lines whatever the galaxy place we tested. We 
therefore deal with a passive object.
NED02 shows strong emission lines only in the galaxy center. This shows that no significant induced 
star-formation (potentially due to gravitational interactions between NED01 and NED02) is present in the galaxy outskirts. 
The central emission lines are most probably due to a central active object, probably a Seyfert.

We also obtained high resolution optical spectroscopy covering a short wavelength range around the H$\alpha$/[NII] domain. We then derived a robust central velocity dispersion that we could translate into a central black hole mass of a few 10$^5$ M$_{\odot}$. This estimate was smaller, but not inconsistent with the one we derived using the optical variability. Given our uncertainties, we therefore conclude that the central black hole has a mass of the order of $\sim$10$^6$ M$_{\odot}$. 
 
 We did not detect short-term variability in the optical ZTF light curves. However, we found a significant long-term stochastic variability in the $g$-band that can be well described by the damped random walk model with a best-fitted characteristic damping timescale of $\tau_{\mathrm{DRW}}=30_{-12}^{+28}$ days.
 
Finally, we have shown that the CGCG 077-102 galaxy pair is deeply embedded within the Abell 2063 cluster potential. Only 5\% of
the cluster galaxies are more linked with the cluster potential than the CGCG 077-102 pair.

As noted in the introduction, CGCG 077-102 is comparable to the J085953.33+131055.3 pair discussed in \cite{2019ApJ...875..117P}.  Unlike the J085953.33+131055.3 system, however, the X-ray source of CGCG 077-102 is not heavily absorbed. \citeauthor{2019ApJ...875..117P} hypothesized that proximity of a neighboring galaxy produced enhanced accretion on the X-ray emitting nucleus of the other galaxy. Enhanced accretion in their model then led to enhanced column densities of $N_{\rm H}$ of order $10^{23}$~cm$^{-2}$. The X-ray source in the nucleus of NED02 has a much lower best fit to its $N_{\rm H}$ of the order of $10^{20}$~cm$^{-2}$. With the caveat that only projected separation of J085953.33+131055.3 and CGCG 077-102 are similar, an explanation of the lower value of Nh for the CGCG 077-102 system is that ram pressure stripping by the ICM of A2063 has (partially?) removed the source of Nh that would otherwise have been available to produce a high value of $N_{\rm H}$ obscuration. The detection of more such CGCG 077-102 pairs in rich clusters will clarify if ram pressure stripping was the cause of the lower value of $N_{\rm H}$ of the X-ray emitting AGN of the CGCG 077-102 pair.

Our black hole mass estimates are based on the analysis of the optical spectrum and optical flux variation of the central region of NED02, respectively  from a few $10^5$
\Ms to a few times $10^6$ \Ms . Yet another method would be to take the Eddington limit for spherical accretion (e.g. \cite{1979rpa..book.....R,2020ARA&A..58..257G}) with the X-ray luminosity of the black hole as about $1.3 \times 10^{38}$\,\es, and then to use a value of $5 \times 10^{42}$\,\es (based on the typical X-ray luminosity we derived, see Table \ref{tab:fitXspec}) for the X-ray luminosity to infer a lower bound to the mass of the central SMBH, giving a lower limit of about $4 \times 10^{4}$\,\Ms .  The lower end of our mass estimates put the mass at the top of the intermediate black hole mass range (IMBH, \cite{2020ARA&A..58..257G}). Therefore, we may have found an IMBH. However, the mass estimate based on the optical spectrum lies just above that $10^5$\,\Ms\ value that is at the low end of the extrapolation of the M-$\sigma$ extrapolation discussed by \cite{2020ARA&A..58..257G} who state they are confident central SMBHs extend to 10$^5$\,\Ms.  Furthermore, \cite{2020ARA&A..58..257G} suggest the highest time variability is expected to come from the most compact emitting region, the X-ray emitting region. Therefore, searching for random and quasi-periodic oscillations (QPOs) will be an interesting follow-up project. 
Evidence for QPOs in the X-rays and its relationship to the mass of the central SMBH has been discussed by, for example \cite{2021ApJ...906...92S,2021Agarwal}. See also \cite{2020MNRAS.499.2380S}. Based on those results, and the estimated mass we have made for the  SMBH in NED02, a plausible  range of frequencies we could hope to detect in X-ray observations are about 10$^{-3}$-10$^{-4}$ Hz.  The are low enough to justify the search in further observation with satellites such as \xmm\ and NICER.  

\begin{acknowledgements}
Authors thank the referee for useful comments.

Authors thank F. Durret, P. Theul\'e, V. LeBrun, and I. Marquez for useful discussions. Authors thank J. Schmitt for his great contribution to the MISTRAL building and maintenance.

Funding for the Sloan Digital Sky Survey IV has been provided by the Alfred P. Sloan Foundation, the U.S. 
Department of Energy Office of Science, and the Participating Institutions. SDSS-IV acknowledges
support and resources from the Center for High-Performance Computing at
the University of Utah. The SDSS web site is www.sdss.org.

SDSS-IV is managed by the Astrophysical Research Consortium for the 
Participating Institutions of the SDSS Collaboration including the 
Brazilian Participation Group, the Carnegie Institution for Science, 
Carnegie Mellon University, the Chilean Participation Group, the French Participation Group, Harvard-Smithsonian Center for Astrophysics, 
Instituto de Astrof\'isica de Canarias, The Johns Hopkins University, 
Kavli Institute for the Physics and Mathematics of the Universe (IPMU) / 
University of Tokyo, the Korean Participation Group, Lawrence Berkeley National Laboratory, 
Leibniz Institut f\"ur Astrophysik Potsdam (AIP), 
Max-Planck-Institut f\"ur Astronomie (MPIA Heidelberg), 
Max-Planck-Institut f\"ur Astrophysik (MPA Garching), 
Max-Planck-Institut f\"ur Extraterrestrische Physik (MPE), 
National Astronomical Observatories of China, New Mexico State University, 
New York University, University of Notre Dame, 
Observat\'ario Nacional / MCTI, The Ohio State University, 
Pennsylvania State University, Shanghai Astronomical Observatory, 
United Kingdom Participation Group,
Universidad Nacional Aut\'onoma de M\'exico, University of Arizona, 
University of Colorado Boulder, University of Oxford, University of Portsmouth, 
University of Utah, University of Virginia, University of Washington, University of Wisconsin, 
Vanderbilt University, and Yale University.

The DESI Legacy Imaging Surveys consist of three individual and complementary projects: the Dark Energy Camera Legacy Survey (DECaLS), the Beijing-Arizona Sky Survey (BASS), and the Mayall z-band Legacy Survey (MzLS). DECaLS, BASS and MzLS together include data obtained, respectively, at the Blanco telescope, Cerro Tololo Inter-American Observatory, NSF’s NOIRLab; the Bok telescope, Steward Observatory, University of Arizona; and the Mayall telescope, Kitt Peak National Observatory, NOIRLab. NOIRLab is operated by the Association of Universities for Research in Astronomy (AURA) under a cooperative agreement with the National Science Foundation. Pipeline processing and analyses of the data were supported by NOIRLab and the Lawrence Berkeley National Laboratory (LBNL). Legacy Surveys also uses data products from the Near-Earth Object Wide-field Infrared Survey Explorer (NEOWISE), a project of the Jet Propulsion Laboratory/California Institute of Technology, funded by the National Aeronautics and Space Administration. Legacy Surveys was supported by: the Director, Office of Science, Office of High Energy Physics of the U.S. Department of Energy; the National Energy Research Scientific Computing Center, a DOE Office of Science User Facility; the U.S. National Science Foundation, Division of Astronomical Sciences; the National Astronomical Observatories of China, the Chinese Academy of Sciences and the Chinese National Natural Science Foundation. LBNL is managed by the Regents of the University of California under contract to the U.S. Department of Energy. 

The Siena Galaxy Atlas was made possible by funding support from the U.S. Department of Energy, Office of Science, Office of High Energy Physics under Award Number DE-SC0020086 and from the National Science Foundation under grant AST-1616414.

Based in part on observations made at Observatoire de Haute Provence (CNRS), France,
with GHASP and MISTRAL.

This research has made use of the MISTRAL database, operated at CeSAM (LAM), Marseille, France.

K.P.R. acknowledges financial support of the \textit{Coordenação de Aperfeiçoamento de Pessoal de Nível Superior} (CAPES), Grant No. 88887.694541/2022-00. GBLN acknowledges partial financial support from CNPq grant 303130/2019-9.

\end{acknowledgements}

\textit{Software Used}: SLINEFIT \citep{2018A&A...618A..85S},
ASPIRED \citep{2022ASPC..532..537L}, numpy \citep{Harris2020}, astropy \citep{2022Astropy}, scipy \citep{2020SciPy}, matplotlib \citep{Hunter2007}, corner \citep{2016corner}, emcee \citep{2013ForemanMackey}, celerite \citep{2017celerite}.

\bibliography{report.bib} 
\bibliographystyle{aa}

\newpage
\begin{appendix}
    
\section{CGCG 077-102 NED01 light curves and CGCG 077-102 NED02 light curve after noise correction ($r$-band)} 
\label{app:ned01_lc}

\begin{figure}[ht]
    \centering 
    \includegraphics[width=9.18cm]{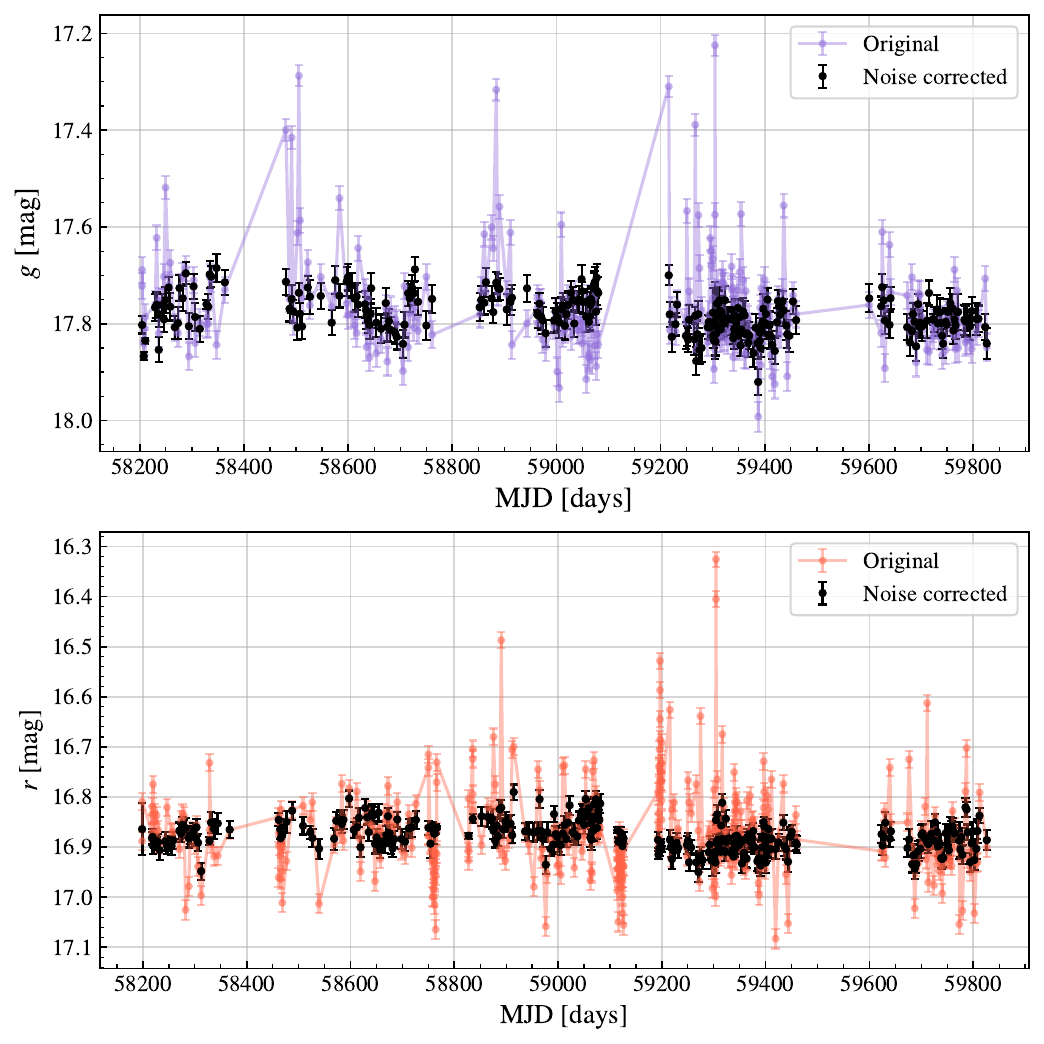}
    \caption[]{Light curves in $g$ (top panel) and $r$ (bottom panel) bands of NED01 before and after the noise correction with PDT and calculation of the intranight observations mean. The purple and orange connected dots represent the original light curves in their respective filters ($g$ and $r$), and the black dots with error bars represent the noise-corrected light curve. Note that the light curve in $r$ band becomes nearly flat after the noise correction, as we expected. However, the light curve in $g$ band shows some periodic-like oscillations that are likely artifacts introduced by the PDT during the detrending process.}
    \label{fig:NED01_detrended}
\end{figure}

\begin{figure}[ht]
    \centering 
    \includegraphics[width=9.18cm]{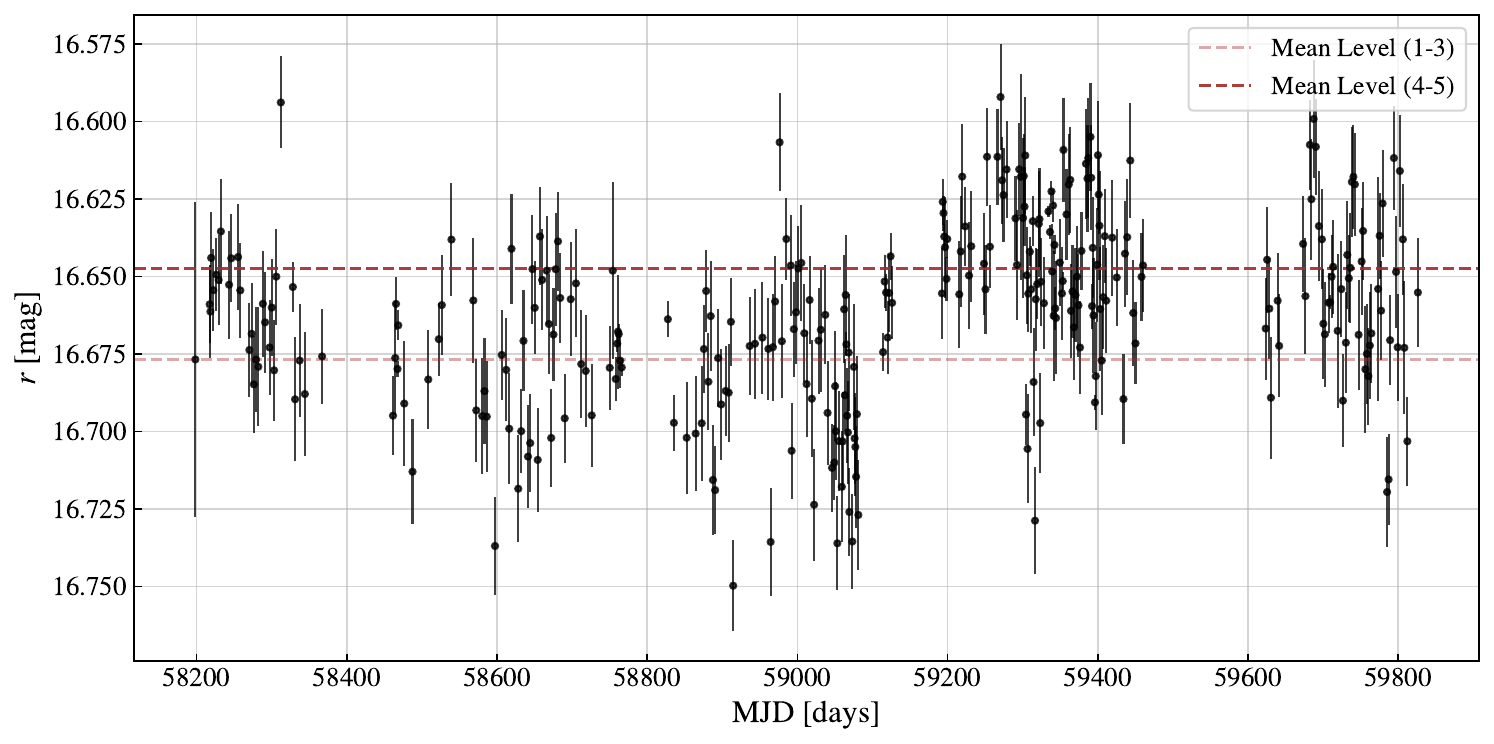}
    \caption[]{Light curve of NED02 in $r$-band (black dots). The dashed lines are the mean levels of the first three (light red) and the last two (dark red) cycles. In this case, we observed a low variability after the noise correction. For this reason, we cannot affirm without any biases whether the light curve presents periodic-like oscillations or artifacts introduced during the detrending process. Furthermore, as discussed in Subsection \ref{subsub:var}, we cannot exclude the possibility of over-subtraction of the signal amplitude, which could have prevented us from analyzing the light curve correctly in this filter. }
    \label{fig:NED02_r_detrended}
\end{figure}

In the framework of section \ref{sec:optical},
we present in Fig. \ref{fig:NED01_detrended} the comparison between the light curves in $g$ and $r$ bands of NED01 before and after the noise correction. Fig.\ref{fig:NED02_r_detrended} gives
the light curve of NED02 galaxy in $r$-band after noise correction.

\section{Lomb-Scargle periodogram}
\label{ap:LSP}

The Lomb-Scargle periodogram (LSP) is a method for periodicity analysis in time-series data \citep{1976Lomb,1982Scargle}. It is more widely used than classical techniques (such as the traditional Fourier methods) in astronomy due to its well-known efficiency in modeling time series with irregularly spaced data and gaps. The LSP is based on the least-squares fit of a sinusoidal model to the data at each frequency, but extensions of the standard LSP (as models that allow multiple Fourier terms and offsets) can also be found in the literature \citep[see][for more details]{2018VanderPlas}.

We employed the \texttt{astropy.timeseries}\footnote{\url{https://docs.astropy.org/en/stable/timeseries/lombscargle.html}} \citep{2013Astropy,2018Astropy,2022Astropy} Python package to implement the LSP method for detecting periodic signals in the light curve. First, we applied a frequency cutoff in the periodogram for the low and high frequencies to avoid the effects of data gaps and white noise that can give us a false estimate of the period. At low frequencies, we cut the frequencies that correspond to periods longer than the interval between the first and last observation of the sample. For the high frequencies, we cut frequencies higher than the mean cadence of the data interval. Here we took advantage of the LSP multiterm extension to get a more flexible model to describe the observed light curve accurately.

\begin{figure*}[ht]
    \centering
    \includegraphics[scale=0.6]{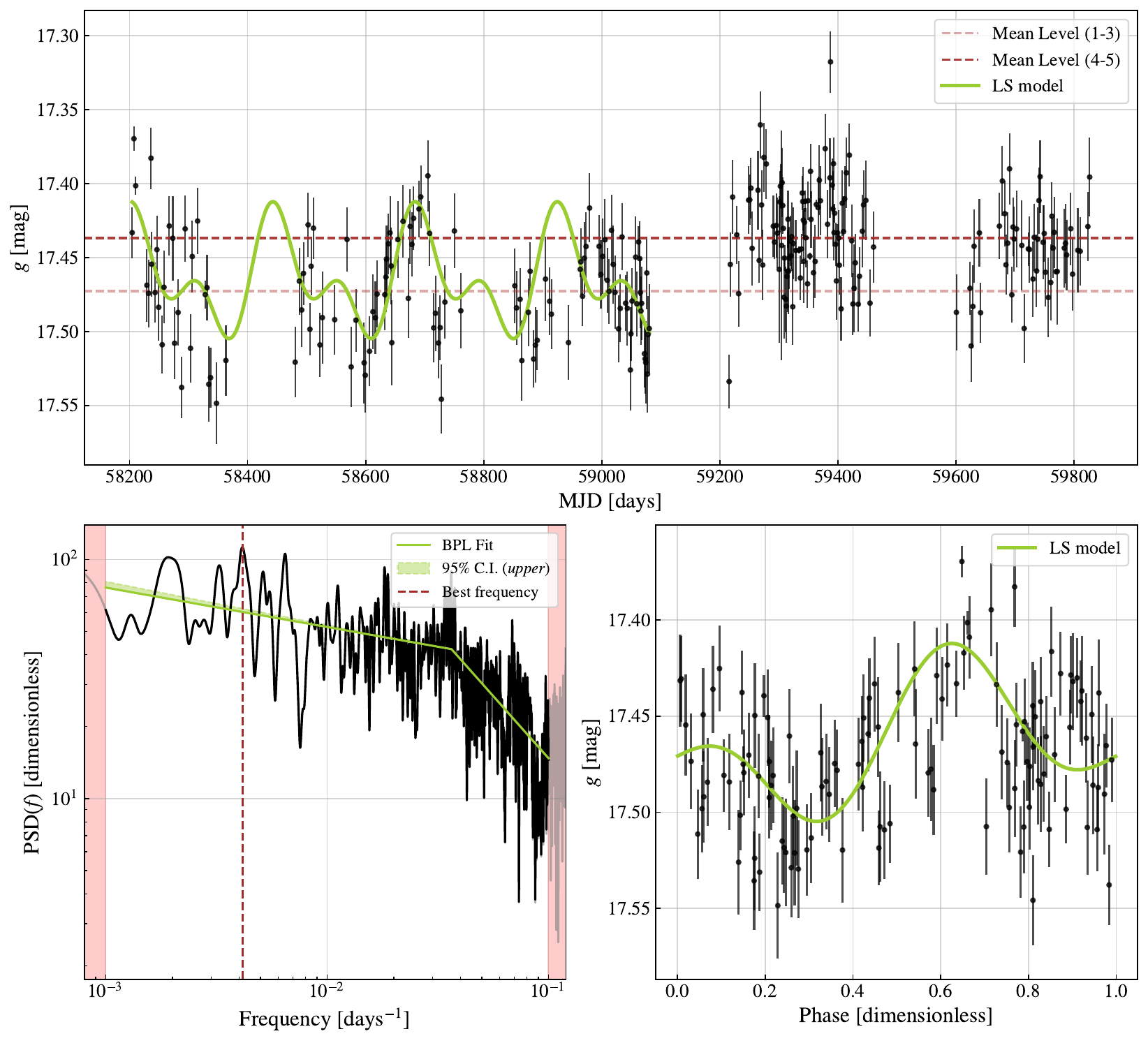}
    \caption[]{LSP analysis. \textit{Top}: the detrended light curve of NED02 galaxy in $g$-band (black dots). The green line represents the best-fitted Lomb-Scargle (LS) model with two Fourier terms on the first three cycles of observations. The dashed lines are the mean levels of the first three (light red) and the last two (dark red) cycles. \textit{Bottom left}: the Lomb-Scargle periodogram. The solid black line represents the power spectrum density (PSD), the light green line is the broken power law (BPL) best fit (a break frequency of $f_{break}=0.036 \pm 0.001$ days$^{-1}$, low-frequency slope of $\alpha_{1}=0.164 \pm 0.007$, and high-frequency slope of $\alpha_{2}=1.04 \pm 0.03$), the light shaded green region is the 95\% upper limit of the confidence interval (C.I.), and the light shaded red region shows the frequency cutoff ($<0.001$ days$^{-1}$ and $>0.1$ days$^{-1}$). The vertical dark red dashed line shows the highest peak above the 95\% upper limit, corresponding to a frequency of $\sim 0.004$ days$^{-1}$. The PSD is dimensionless because the uncertainties were specified within the \texttt{LombScargle}/astropy implementation. \textit{Bottom right}: the phase-folded light curve (black dots) for the first three cycles of observations, with the best-fitted LS model (solid green line). The folding period is $240.9$ days. }
    \label{fig:NED02_LS}
\end{figure*} 

The LSP algorithm was not able to model the entire light curve correctly, likely due to the large data gaps between the last cycles of observations and the different mean levels in two distinct intervals of the light curve, as reported in Subsection \ref{subsub:var} (see also Figure \ref{fig:NED02_LS}). For this reason, we applied the LSP algorithm for the first three cycles of observations, excluding the last two from the analysis. Furthermore, we could not analyze the last two cycles separately because the algorithm fails to model this interval due to the gaps and the low number of observations. 


Although we expected a noisier periodogram due to the multiterm Fourier model, which makes the detection of the correct peak difficult, the computed periodogram for a model with two Fourier terms shows an additional noise that increases toward low frequencies, similar to a shape of a power spectrum density (PSD) dominated by a red/pink noise (PSD $\propto f^{-\alpha}$). For this reason, we fitted a broken power law\footnote{\url{https://docs.astropy.org/en/stable/api/astropy.modeling.powerlaws.BrokenPowerLaw1D.html}} on the PSD to estimate the red/pink noise continuum \citep[see more details about "colored" noise PSDs in][for example]{2005Vaughan,2016Zhu,2017Guo}, and we computed its confidence intervals with the bootstrap method. Here we consider the highest peak above the 95\% upper limit of the confidence interval the most probably correct peak. For a detailed discussion about why measuring the period uncertainties in terms of error bars is not recommended, see \cite{2018VanderPlas}.

The best-fitted Lomb-Scargle (LS) model with two Fourier terms on the first three cycles of observations, the power spectrum density, and the phase-folded light curve are shown in Figure \ref{fig:NED02_LS}. Doubtless, the Lomb-Scargle model is not flexible enough to describe the entire NED02 light curve, which requires a more complex model to fulfill this task, such as stochastic models. Also, it is well known that red noise signal can be confused with true periodic oscillations, leading us to detect false periodicities in AGNs \citep[e.g.,][]{2016Vaughan}. Therefore, this method cannot confirm the presence of a real periodic signal in the NED02 light curve.

\end{appendix}

\end{document}